\documentclass[showpacs,amssymb,preprint,preprintnumbers,nofootinbib,superscriptaddress]{revtex4}
\usepackage[dvips]{graphicx}
\usepackage{amsmath}
\usepackage{graphicx}
\usepackage{amsfonts}
\usepackage{amssymb}
\usepackage{epsfig}
\usepackage{color}

\graphicspath{{Figures/}}
\usepackage{latexsym}
\usepackage{url,hyperref}
\usepackage{bm}
\usepackage{textcomp}
\usepackage{color}
\usepackage{bbm}
\usepackage{slashed}
\usepackage{caption}
\usepackage{epstopdf}
\usepackage{subcaption}
\usepackage{placeins}
\usepackage{diagbox}

\begin{document}
\title{\bf Greybody factor for massive fermion emitted by a black hole in dRGT massive gravity theory}

\author{P. Boonserm$^{1,4}$\hspace{1mm}, C. H. Chen$^{2}$, T. Ngampitipan$^{3}$ and P. Wongjun$^{2,4}$\\
$^{1}$  {\small {\em Department of Mathematics and Computer Science, Faculty of Science,}}\\
{\small {\em Chulalongkorn University, Bangkok 10330, Thailand}}\\
$^{2}$  {\small {\em The institute for fundamental study,}}\\
{\small {\em Naresuan University, Phitsanulok 65000, Thailand}}\\
$^{3}$  {\small {\em Physics Program, Faculty of Science, Chandrakasem Rajabhat University,}}\\
{\small {\em Bangkok 10900, Thailand}}\\
$^{4}$  {\small {\em Thailand Center of Excellence in Physics,}}\\
{\small {\em Ministry of Higher Education, Science, Research and Innovation, Bangkok 10400, Thailand}}\\
}
\maketitle

\section*{Abstract}
The greybody factor of the massive Dirac field around the black hole in the dRGT massive gravity theory is investigated using the rigorous bound and the WKB approximation methods. In both methods, the greybody factor significantly depends on the shape of the potential. If the potential is smaller, there is more probability for the Dirac field to transmit through the black hole, therefore, the greybody factor is higher. Moreover, there exists a critical mass of the Dirac field such that the greybody factor is maximum. By comparing the results from these two methods, we argue that it is useful to use the rigorous bound method for the low potential cases, while using the WKB approximation method for the high potential cases.

\section{Introduction}

One of the important predictions of General Relativity (GR) is the possibility of existence of a mysterious object known as black holes. Even though it is not believed to exist in the real world at the beginning era after GR was proposed, Roger Penrose proved that black holes really can form using ingenious mathematical methods \cite{Penrose:1964wq}. Together with observational data of Sagittarius A* at the centre of our galaxy, it suggests that black holes really can exist. Recently, by using the advancing techniques in radio interferometry, the first image of a black hole has been detected by The Event Horizon Telescope (EHT) \cite{Akiyama:2019cqa}. The information of the image is used to constraint the properties of black holes \cite{Garofalo:2020ajg,Vincent:2020dij,Dokuchaev:2020rye}, and also constraint how the modified gravity theories can be deviated from GR \cite{Stepanian:2021vvk}. Moreover, the direct detection of gravitational wave is one of the strong evidences of the existence of black holes recently \cite{TheLIGOScientific:2016src}. The information of the detection also provides constraint on the modified gravity theory, for example the speed of gravitational wave \cite{Cornish:2017jml}. These may provide the reason for why the study of black holes receives much attention nowadays.

One of the most important characteristic behaviour is that black holes behave as a thermal system. Particularly, black holes carry entropy and can emit a type of radiation called the Hawking radiation \cite{Hawking:1974sw,Hawking:1976de}. As a result, at the event horizon, the spectrum of the radiations from black holes is the same as that of the black-body spectrum. Since the spacetime around the black hole is curved, the spectrum emitted from the black hole is significantly modified. In this sense, the spacetime curvature can act as a potential barrier which allows some of the radiation to transmit and reflect as found in similar situation in quantum mechanics. As a result, the greybody factor is defined in order to take into account the transmission amplitude of the radiation from the black hole. On the other hand, it can be viewed as the probability for a wave coming from infinity to be absorbed by the black hole, which is sometimes referred to as the rate of absorption probability.

The greybody factors from various kinds of spacetime geometry have been intensively investigated by various methods. Using a similar strategy as with quantum mechanics, one can find the transmission amplitude by finding solutions in asymptotic regions and then matching the solutions at the boundaries \cite{Fernando:2004ay,Ahmed:2016lou,Sharif:2019yiy,Sharif:2020hyz}. Usually, the solutions are written in terms of special functions, which make it difficult to analyze the behaviour of the spectrum analytically. One of the possibilities which is intensively investigated in literature is that of using WKB approximation \cite{Parikh:1999mf,Cho:2004wj,Konoplya:2010kv,Dey:2018cws,Konoplya:2019ppy,Konoplya:2019hlu,Devi:2020uac}. It provides a good approximation for a simple form of the spacetime geometries, which then requires the higher potential, or in the other words, requires high multipole. The other way to investigate the greybody factor is to consider the bound of the greybody factor instead of the exact one \cite{Visser:1998ke,Boonserm:2008qf,Boonserm:2009zba,Boonserm:2017qcq,Boonserm:2019mon,Barman:2019vst,Chowdhury:2020bdi,Kanzi:2020cyv}. This method allows us to study the behaviour of the greybody factor analytically.

There have been many attempts to modify GR due to a cosmological aspect as the universe expands with an acceleration \cite{Supernova,Supernova2}. Such modified gravity theories must be reduced to GR at a local scale in order to satisfy the concrete observations. In this context, the black hole solution will obtain additional corrections due to the modifications, and then the properties of the black hole, such as the greybody factor, may significantly be modified. In the present work, we consider de Rham-Gabadadze and Tolley (dRGT) massive gravity theory \cite{deRham:2010ik,deRham:2010kj} where the black hole solutions are proposed in \cite{Berezhiani:2011mt,Brito:2013xaa,Volkov:2013roa,Cai:2012db,Babichev:2014fka,Babichev:2015xha,Hu:2016hpm,Cai:2014znn,Ghosh:2015cva}. The dRGT massive gravity theory is a modified gravity theory such that the systematic mass terms are included into GR in order to eliminate the additional ghost degree of freedom (see \cite{Hinterbichler,deRham:2014zqa} for review). One of the key points of the dRGT massive gravity is that the Struckelberg fields are introduced via the reference/fiducial metric to restore the diffeomorphism invariance.

In the context of Hawking radiation, there are various kinds of fluctuation fields treated as the radiation while the massless case in dRGT massive gravity has been investigated recently \cite{Hou:2020yni}. Therefore, studies of greybody factors around the black hole depends on particular fields. For the black hole in dRGT massive gravity theory, the greybody factor from scalar field has been investigated \cite{Boonserm:2017qcq,Kanzi:2020cyv}, while the part from the Dirac fermion field have not been investigated yet. In actually, the matter fields around the black hole is supposed to be the fermion fields, therefore, it becomes worthwhile to investigate the fluctuations around the black hole as the fermion field. In the current paper, we aim to investigate the greybody factor in dRGT massive gravity theory due to the Dirac fermion field.

The paper is organized as follows. We first review the basic knowledge of the dRGT massive gravity as well as the black hole solutions in  Section \ref{back}. Moreover, the equations of motion related to the Dirac field around the dRGT black hole is also reviewed in this section. Using these equations of motion with specific   potential, the rigorous bound of the greybody factor is investigated in Section \ref{bound}. It is found that the greybody factor significantly depends on the shape of the potential. If the potential is smaller, there is more probability for the Dirac field to transmit through the black hole, therefore, the greybody factor is higher. Moreover, there exists a critical mass of the Dirac field such that the greybody factor is maximum. We also found that for large multipole, the bound is much lower than the exact value, therefore, this method may not be useful for the large multipole $\lambda$ corresponding to the high potential case. We also investigated the greybody factor using the WKB approximation in Section \ref{WKB}. The main results agree with the rigorous bound method. However, the WKB method does not work well for low multipole $\lambda$, which corresponds to the low potential cases. Finally, we summarize the results in Section \ref{con}.

\section{Background}\label{back}
In this section, we will review the basic knowledge about dRGT massive gravity and their black hole solutions, as well as the Dirac field around the dRGT black hole.

\subsection{dRGT massive gravity theory}

The dRGT massive gravity theory is one of the viable models of massive gravity theories proposed by de Rham, Gabadaze and Tolley  \cite{deRham:2010ik,deRham:2010kj}. The action of the theory is the Einstein-Hilbert action, added with suitable mass terms as follows
\begin{eqnarray}\label{action}
 S = \int d^4x \sqrt{-g}\; \frac{1}{2} \left[ R +m_g^2\,\, {\cal U}(g, \phi^a)\right],
\end{eqnarray}
where $R$ is the Ricci scalar corresponding to the kinetic part of the gravitational field and ${m_g^2 \cal U}$ corresponds to a potential part with graviton mass $m_g$. In a four-dimensional spacetime, the potential ${\cal U}$  can be expressed as
\begin{eqnarray}\label{potential}
	{\cal U}(g, \phi^a) = {\cal U}_2 + \alpha_3{\cal U}_3 +\alpha_4{\cal U}_4 ,
\end{eqnarray}
where $\alpha_3$ and $\alpha_4$ are dimensionless free parameters of the theory. The potential ${\cal U}_2$, ${\cal U}_3$ and ${\cal U}_4$ can be written in terms of the physical metric tensor $g_{\mu\nu}$ and fiducial metric tensor $f_{\mu\nu}$ as
\begin{eqnarray}
	{\cal U}_2&\equiv&[{\cal K}]^2-[{\cal K}^2] ,\\
	{\cal U}_3&\equiv&[{\cal K}]^3-3[{\cal K}][{\cal K}^2]+2[{\cal K}^3] ,\\
	{\cal U}_4&\equiv&[{\cal K}]^4-6[{\cal K}]^2[{\cal K}^2]+8[{\cal K}][{\cal
K}^3]+3[{\cal K}^2]^2-6[{\cal K}^4],
\end{eqnarray}
where
\begin{eqnarray}
	{\cal K}^\mu_{\,\,\,\nu}=\delta^\mu_\nu-\sqrt{g^{\mu\sigma}f_{ab}\partial_\sigma\phi^a\partial_\nu\phi^b}. \label{K-tensor}
\end{eqnarray}
Note that the rectangular brackets denote the traces, namely $[{\cal K}]={\cal K}^\mu_{\,\,\,\mu}$ and $[{\cal K}^n]=({\cal K}^n)^\mu_{\,\,\,\mu}$. The four scalar fields $\phi^a$ are St\"uckelberg fields introduced in order to restore the general covariance of the theory.

The equations of motion can be obtained by varying the above action as follows
\begin{eqnarray}
	G_{\mu\nu} +m_g^2 X_{\mu\nu} = 0, \label{modEFE}
\end{eqnarray}
where the tensor $X_{\mu\nu}$ is the result from varying the potential term ${\cal U}$ with respect to $g_{\mu\nu}$ expressed as
\begin{eqnarray}
	X_{\mu\nu} &=& {\cal K}_ {\mu\nu} -{\cal K}g_ {\mu\nu} -\alpha\left({\cal K}^2_{\mu\nu}-{\cal K}{\cal K}_{\mu\nu} +\frac{{\cal U}_2}{2}g_{\mu\nu}\right) 
 	+3\beta\left( {\cal K}^3_{\mu\nu} -{\cal K}{\cal K}^2_{\mu\nu} +\frac{{\cal U}_2}{2}{\cal K}_{\mu\nu} - \frac{{\cal U}_3}{6}g_{\mu\nu} \right). \,\,\,\,\,\,\label{effemt}\nonumber\\
\end{eqnarray}
Note that we have reparameterized the model parameters as follows
\begin{eqnarray}\label{alphabeta}
	\alpha_3 = \frac{\alpha-1}{3}~,~~~\alpha_4=\frac{\beta}{4}+\frac{1-\alpha}{12}.
\end{eqnarray}
Since the potential terms are covariantly constructed, the tensor $X_{\mu\nu}$ obeys the covariant divergence as follows
\begin{eqnarray}\label{BiEoM}
	\nabla^\mu X_{\mu\nu} = 0,
\end{eqnarray}
where $\nabla^\mu$ denotes the covariant derivative, which is compatible with $g_{\mu\nu}$. Note that, this constraint equation is also obtained by varying the action with respect to the fiducial metric, which also satisfies the Bianchi identities.

\subsection{Black hole solutions}

In this subsection, we review black hole solutions in dRGT massive gravity called dRGT black hole solution using the following Ref. \cite{Ghosh:2015cva}. In order to solve the solution for the field equation in Eq.~\eqref{modEFE}, one needs to specify the form of the fiducial metric. In this consideration, it is convenient to choose the form of the fiducial metric as

\begin{eqnarray}\label{fiducial metric}
f_{\mu\nu}&=&\text{diag}(0,0,c^2  ,c^2 \sin^2\theta), \label{fmetric}
\end{eqnarray}
where $c$ is a constant. By substituting this fiducial metric, one of the static and spherically symmetric solutions of the physical metric can be obtained as
\begin{eqnarray}
\text{d}s^2&=&-f\text{d}t^2+\frac{1}{f}\text{d}r^2+r^2(\text{d}\theta^2+\sin^2\theta\text{d}\phi^2),\\
	f(\tilde{r}) &=& 1 - \frac{2\tilde{M}}{\tilde{r}} + \alpha_g \left(c_2\tilde{r}^2 -c_1 \tilde{r}+c_0\right),\label{fdRGT}
\end{eqnarray}
where we rescale the radial coordinate $r= c \tilde{r}$ as well as the model parameters as follows
\begin{eqnarray}
	\tilde{M}&=& \frac{M}{c},\,\,\,\,\,\,
	\alpha_g = m^2_gc^2,\,\,\,\,\,\,
	c_0 = \alpha+3\beta,\,\,\,\,\,\,
	c_1=1 +2\alpha +3\beta,\,\,\,\,\,\,
	c_2 = 1+\alpha+\beta.
\end{eqnarray}
Note that the scale of $c$ takes place at $\tilde{M} \sim \alpha_g$ and then corresponds to the Vainshtein radius
\begin{eqnarray}
	c = r_V = \left(\frac{M}{m_g^2}\right)^{1/3}.
\end{eqnarray}
The theory in which $r < r_V$ will approach GR, while the theory in which $r > r_V$, the modification of GR will be active. Note that detailed calculation to obtain the solution can be found in \cite{Ghosh:2015cva}. This solution contains various signatures of other well-known black hole solutions found in literature. By setting $\alpha_{g} = 0$, the Schwarzschild (Sch) solution is recovered. For the very large scale limit with $\alpha_g > 0$, the solution becomes the Schwarzschild-de-Sitter (Sch-dS) solution for $c_2 < 0$ and  becomes the Schwarzschild-anti-de-Sitter (Sch-AdS) solution for $c_2 > 0$. Moreover, the last term ($c_0$ term) in Eq.~\eqref{fdRGT} corresponds to the global monopole term, which naturally emerges from the graviton mass. Finally, the linear term ($c_1$) is a signature term of this solution, distinguished from other solutions found in literature.

\begin{figure}[h!]
\begin{center}
\includegraphics[scale=0.50]{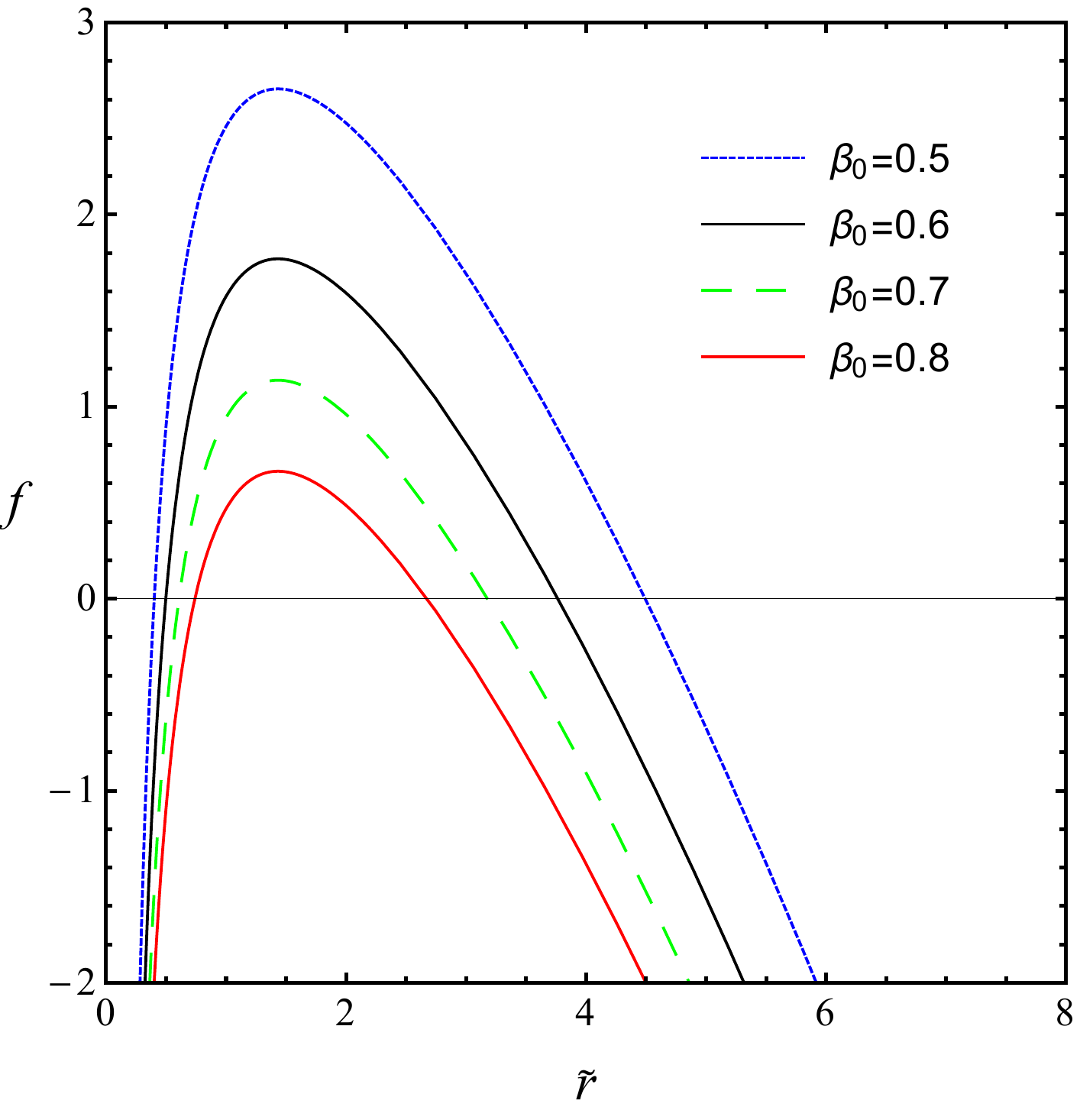}
\end{center}
{\caption{This figure shows the horizon structure of the dS solution for specific values of the parameters being $\alpha_g = M=1, c_2 = \frac{-0.2}{3}$.}\label{fig:dSAdShorizon}}
\end{figure}

From this solution, it is possible to have three horizons for the asymptotic Sch-AdS solution. Therefore, we will restrict our consideration for the asymptotic Sch-dS solution, which has at most two horizons. In order to see the structure of the black hole horizon clearly, let us consider a subclass of parameters, specifying the parameter as follows \cite{Boonserm:2017qcq}
\begin{eqnarray}\label{c1c0}
	c_1 = 3 (4c^2_2)^{1/3},\quad c_0 = \frac{9}{\sqrt{3}} \frac{\left(2|c_2|\right)^{1/3}}{\beta_m} - \frac{1}{\alpha_g}.
\end{eqnarray}
By choosing this reparametrization, it allows us to characterize the existence of the horizons by $0< \beta_m <1$ and the strength of the graviton mass by parameter $c_2$. The maximum point of $f$ occurs at $\tilde{r}_{max} = (- 2 c_2)^{-1/3}$ where the maximum value of $f$ can be written as
\begin{equation}
f(\tilde{r}_{max}) =  \frac{9\alpha_g \left(-2c_2\right)^{1/3}}{\sqrt{3}\beta_m}(1-\beta_m).
\end{equation}
By using this parameter setting, two real positive horizons can be solved and then written in terms of $\beta_m$ and $c_2$ as
\begin{eqnarray}
\tilde{r}_{H} &=& \frac{2}{\left(-2c_2\right)^{1/3}}\left[X^{1/2}\cos\left(\frac{1}{3}\sec^{-1}Y \right) - 1\right],\\
\tilde{R}_{H} &=& \frac{-2}{\left(-2c_2\right)^{1/3}}\left[X^{1/2}\cos\left(\frac{1}{3}\sec^{-1} Y+ \frac{\pi}{3}\right) + 1\right] ,
\end{eqnarray}
where $r_H$ denotes the black hole horizon, $R_H$ denotes the cosmic horizon, and
\begin{eqnarray}
X= \frac{2\sqrt{3}}{\beta_m} + 4,\quad Y=\frac{X^{3/2}\beta_m}{2(5\beta_m+3\sqrt{3})}.
\end{eqnarray}
The structure of the horizons parameterized by $\beta_m$ can be seen explicitly in Fig. \ref{fig:dSAdShorizon}.

\subsection{Equations of motion of the massive Dirac field}\label{eom}
In this section, we review the equation of motion of the Dirac field present in the background of black holes using the following Ref. \cite{Chen:2019kaq}. For brevity, we will omit ``tilde" in radial coordinate as well as the black hole mass parameter. As a result, the general form of the metric can be written in the form
\begin{eqnarray}
	\text{d}s^2=-f(r)\text{d}t^2+\frac{1}{f(r)}\text{d}r^2+r^2(\text{d}\theta^2+\sin^2\theta\text{d}\phi^2).
\end{eqnarray}
For the spin-half fields in curved spacetime, it is convenient to use vielbein formalism where the vielbein can be defined as
\begin{eqnarray}
	e^\mu_{\,\,\,\hat{\alpha}}&=&\text{diag}\left(\frac{1}{\sqrt{f}}, \sqrt{f}, \frac{1}{r}, \frac{1}{r\sin\theta}\right).
\end{eqnarray}
The massive Dirac field $\Psi$ of mass $m$ obeys the equation of motion
\begin{eqnarray}
	\Big[\gamma^\mu(\partial_\mu+\Gamma_\mu)+m\Big]\Psi=0.\label{Dirac eq}
\end{eqnarray}
Here, $\gamma^\mu$ is the $4\times4$ Dirac gamma matrix and $\Gamma_\mu$ is the spin connection, which can be expressed in terms of the Christoffel symbol $\Gamma^\rho_{\mu\nu}$ as follows
\begin{eqnarray}
	\Gamma_\mu&=&\frac{1}{2}\,\omega_{\mu\hat{\alpha}\hat{\beta}}\,\Sigma^{\hat{\alpha}\hat{\beta}}, \quad
	\omega_{\mu\hat{\alpha}\hat{\beta}}=e^\rho_{\,\,\,\hat{\alpha}}(\partial_\mu e_{\rho\hat{\beta}}-\Gamma^\sigma_{\mu\rho}e_{\sigma\hat{\beta}}),\quad
	\Sigma^{\hat{\alpha}\hat{\beta}}=\frac{1}{4}[\gamma^{\hat{\alpha}},\gamma^{\hat{\beta}}]
\end{eqnarray}

In this work, we use the representation of the Dirac gamma matrices, $\gamma^{\hat{\alpha}}$, in terms of the Pauli spin matrices $\sigma^i$ as
\begin{eqnarray}
	\gamma^{\hat{0}}=i\sigma^3\otimes\mathbbm{1},\hspace{1cm}
	\gamma^{\hat{1}}=\sigma^2\otimes\mathbbm{1},\hspace{1cm}
	\gamma^{\hat{2}}=\sigma^1\otimes\sigma^1,\hspace{1cm}
	\gamma^{\hat{3}}=-\sigma^1\otimes\sigma^2,
\end{eqnarray}
By using this setting, the Dirac equation in Eq. \eqref{Dirac eq} can be reexpressed as
\begin{eqnarray}
	\left[
	\frac{1}{\sqrt{f}}\Big\{(i\sigma^3\otimes\mathbbm{1})\partial_t+\frac{f'}{4}(\sigma^2\otimes\mathbbm{1})\Big\}
	+\sqrt{f}(\sigma^2\otimes\mathbbm{1})\partial_r
	+\frac{1}{r}(\sigma^1\otimes\sigma^1)\partial_\theta
	\right.\hspace{1cm}&&\nonumber\\
	\left.+\frac{\sqrt{f}}{2r}(\sigma^2\otimes\mathbbm{1})
	-\frac{1}{r\sin\theta}(\sigma^1\otimes\sigma^2)\partial_\phi
	+\frac{\sqrt{f}}{2r}(\sigma^2\otimes\mathbbm{1})
	+\frac{\cot\theta}{2r}(\sigma^1\otimes\sigma^1)	
	\right]&&\Psi=0.\,\,\,\label{Dirac eqf}
\end{eqnarray}
where prime denotes the derivative with respect to $r$. In order to solve this equation, one can use the separation variable method as
\begin{eqnarray}
	\Psi(t,r,\theta,\phi)=\left(\begin{array}{c}iA(r)\\B(r)\end{array}\right)e^{-i\omega t}\otimes\Theta(\theta,\phi),
\end{eqnarray}
where $A$ and $B$ are the radial functions and $\omega$ is the angular frequency of the solution. By substituting this solution into Eq. \eqref{Dirac eqf}, the equation for the redial part can be written as
\begin{eqnarray}
	\left[\left(f\partial_r+\frac{f'}{4}+\frac{f}{r}\right)\sigma^2
	+\frac{i\lambda\sqrt{f}}{r}\sigma^1\right]
	\left(\begin{array}{c}iA\\B\end{array}\right)
	&=&
	-\left[\omega\sigma^3+m\sqrt{f}\mathbbm{1}\right]
	\left(\begin{array}{c}iA\\B\end{array}\right),\label{Dirac eq2}
\end{eqnarray}
where $ \lambda=l+1 =\pm1,\pm2,\pm3,\hdots$ are the eigenvalues for the angular part that obeys the following equation
\begin{eqnarray}
	\left(\sigma^1\partial_\theta-\frac{\sigma^2}{\sin\theta}\partial_\phi+\frac{\cot\theta}{2}\sigma^1\right)\Theta=i\lambda\Theta,
\end{eqnarray}
By using further radial functions, $F(r)$ and $G(r)$ (for more detailed calculations, see Ref. \cite{Chen:2019kaq}), one can decouple the redial equations as follows
\begin{eqnarray}\label{RE}
	\left(-\partial^2_{r_{*}}+V_+\right)F &=&\omega^2 F,\\
	\left(-\partial^2_{r_{*}}+V_-\right)G &=&\omega^2 G,
\end{eqnarray}
where $r_{*}$ is the tortoise coordinate defined by
\begin{eqnarray}\label{ttcoord}
	\frac{dr_{*}}{dr} = \frac{f}{b},
	\quad
	b=1+\frac{fm\lambda}{2\omega(m^2r^2+\lambda^2)}.
\end{eqnarray}
The potential $V_\pm$ can be expressed as
\begin{eqnarray}
	V_\pm = \pm\frac{dW}{dr_{*}} + W^{2},\quad W= \frac{a}{b}, \quad a=\frac{\sqrt{f}}{r}\sqrt{\lambda^2+m^2r^2}.\label{potential}
\end{eqnarray}
where $W$ can be expressed as
\begin{eqnarray}
 W= \frac{a}{b} = \frac{\left(\sqrt{f(r)}/r\right)\sqrt{\lambda^{2} + m^{2}r^{2}}}{1 + \left(1/2\omega\right)f(r)\left[\lambda m/\left(\lambda^{2} + m^{2}r^{2}\right)\right]} .\label{Eq:W}
\end{eqnarray}
One can see that these equations are in the form of the Schr\"{o}dinger-like equations. Moreover, one can see that the potential depends on both the mass and the energy of the Dirac field. This is a crucial property of this potential that is distinguished from the particular ones in quantum mechanics. We will see later, this makes it much more difficult to analyze the resulting greybody factor. Note that the tortoise coordinate $r_{*}$ can be written as
\begin{equation}\label{ttcoor}
r_{*}=\kappa_{r_{H}}\ln\left|\frac{r}{r_{H}}-1\right|-\kappa_{R_{H}}\ln\left|\frac{r}{R_{H}}-1\right|+\kappa_{r_{(-)}}\ln\left|\frac{r}{r_{(-)}}-1\right|+\frac{1}{2\omega}\arctan\left(\frac{mr}{\lambda}\right),
\end{equation}
where $r_{(-)}$ is a negative root for $f=0$. $\kappa_{i}$ are constants  defined on the horizon as
\begin{equation}
\kappa_{i}=\left(\left|\frac{df(r)}{dr}\right|_{r=j}\right)^{-1},
\end{equation}
where $j$ denotes $r_{H}$, $R_{H}$ and $r_{-}$. One can check that when $r\rightarrow r_{H}$, $r_{*}\rightarrow -\infty$ and $r\rightarrow R_{H}$, $r_{*}\rightarrow \infty$. It is convenient to write down the explicit form of the effective potentials as
\begin{eqnarray}\label{veff}
V_{\pm}&=&\frac{\sqrt{f}\left(\lambda^{2}+m^{2}r^{2}\right)^{3/2}}{\left(\lambda^{2}+m^{2}r^{2}+\frac{\lambda m}{2\omega}f\right)^{2}}\left[\frac{\sqrt{f}}{r^{2}}\left(\lambda^{2}+m^{2}r^{2}\right)^{3/2}\pm\left(\frac{f'}{2r}-\frac{f}{r^{2}}\right)\left(\lambda^{2}+m^{2}r^{2}\right)\pm3m^{2}f\right]\nonumber\\
&&\mp\frac{f^{3/2}\left(\lambda^{2}+m^{2}r^{2}\right)^{5/2}}{r\left(\lambda^{2}+m^{2}r^{2}+\frac{\lambda m}{2\omega}f\right)^{3}}\left[2m^{2}r+\frac{\lambda m}{2\omega}f'\right].
\end{eqnarray}
As a remark, $V_{\pm}$ are known as the super-symmetric partner potentials which are isospectra, as such in the later sections, we mainly focus on the studies of the $V_{-}$ potential.

\section{The rigorous bounds on the greybody factors}\label{bound}
In this section, we will investigate the greybody factor using the rigorous bounds. By using this method, it allows us to qualitatively analyze the results. As a result, the effect of the potential on the greybody factor can be determined. The rigorous bounds on the greybody factors are given by
\begin{equation}
T \geq \text{sech}^{2}\left(\int_{-\infty}^{\infty}\vartheta dr_{*}\right),
\end{equation}
where
\begin{equation}
\vartheta = \frac{\sqrt{[h'(r_{*})]^{2} + \left[\omega^{2} - V(r_{*}) - h^{2}(r_{*})\right]^{2}}}{2h(r_{*})}
\end{equation}
and $h(r_{*})$ is a positive function satisfying $h(-\infty) = h(\infty) = \omega$. See Ref. \cite{Visser:1998ke} for more details. We select $h = \omega$. Therefore,
\begin{equation}
T \geq \text{sech}^{2}\left(\frac{1}{2\omega}\int_{-\infty}^{\infty}|V|dr_{*}\right).\label{T}
\end{equation}
Substituting potential from Eq. (\ref{potential}) in the above equation, we obtain
\begin{eqnarray}
T &\geq& \text{sech}^{2}\left(\frac{1}{2\omega}\int_{-\infty}^{\infty}\left|\pm\frac{dW}{dr_{*}} + W^{2}\right|dr_{*}\right), \nonumber \\
  &\geq& \text{sech}^{2}\left(\frac{1}{2\omega}\int_{-\infty}^{\infty}\left|\pm\frac{dW}{dr_{*}}\right|dr_{*} + \frac{1}{2\omega}\int_{-\infty}^{\infty}\left|W^{2}\right|dr_{*}\right) = T_{b}.\label{TT}
\end{eqnarray}
Let us consider separately the first and the second integrals in Eq. \eqref{TT}. For the first integral, we have
\begin{equation}
\int_{-\infty}^{\infty}\left|\pm\frac{dW}{dr_{*}}\right|dr_{*} = \left.W\right|_{r = r_{H}}^{r = R_{H}} = 0.
\end{equation}
The integral vanishes since the function $W$ in Eq. \eqref{Eq:W} is proportional to $\sqrt{f}$ and $f$ vanishes at the horizons. As a result, the rigorous bound can be obtained using only the second integral. By evaluating the integral in Eq. \eqref{TT}, it can be expressed as
\begin{eqnarray}
\int_{-\infty}^{\infty}\left|W^{2}\right|dr_{*} &=& \int_{-\infty}^{\infty}\frac{\left|f(r)\right|}{r^{2}}\frac{\lambda^{2} + m^{2}r^{2}}{\left[1 + \left(1/2\omega\right)\left|f(r)\right|\left[\lambda m/\left(\lambda^{2} + m^{2}r^{2}\right)\right]\right]^{2}}dr_{*}\nonumber\\
      &=& \int_{r_{H}}^{R_{H}}\frac{1}{r^{2}}\frac{\lambda^{2} + m^{2}r^{2}}{1 + \left(1/2\omega\right)\left|f(r)\right|\left[\lambda m/\left(\lambda^{2} + m^{2}r^{2}\right)\right]}dr\nonumber\\
      &=& \int_{r_{H}}^{R_{H}}\frac{1}{r^{2}}\frac{\left(\lambda^{2} + m^{2}r^{2}\right)^{2}}{\lambda^{2} + m^{2}r^{2} + \left(\lambda m/2\omega\right)\left|f(r)\right|}dr.\label{Ab}
\end{eqnarray}
The results of this formulation is significantly different between the massless and the massive cases. Therefore, we separate our consideration case by case.

\subsection{massless fermion}

\begin{figure}[h!]
\begin{center}
\begin{subfigure}{0.4\textwidth}
\includegraphics[width=\textwidth]{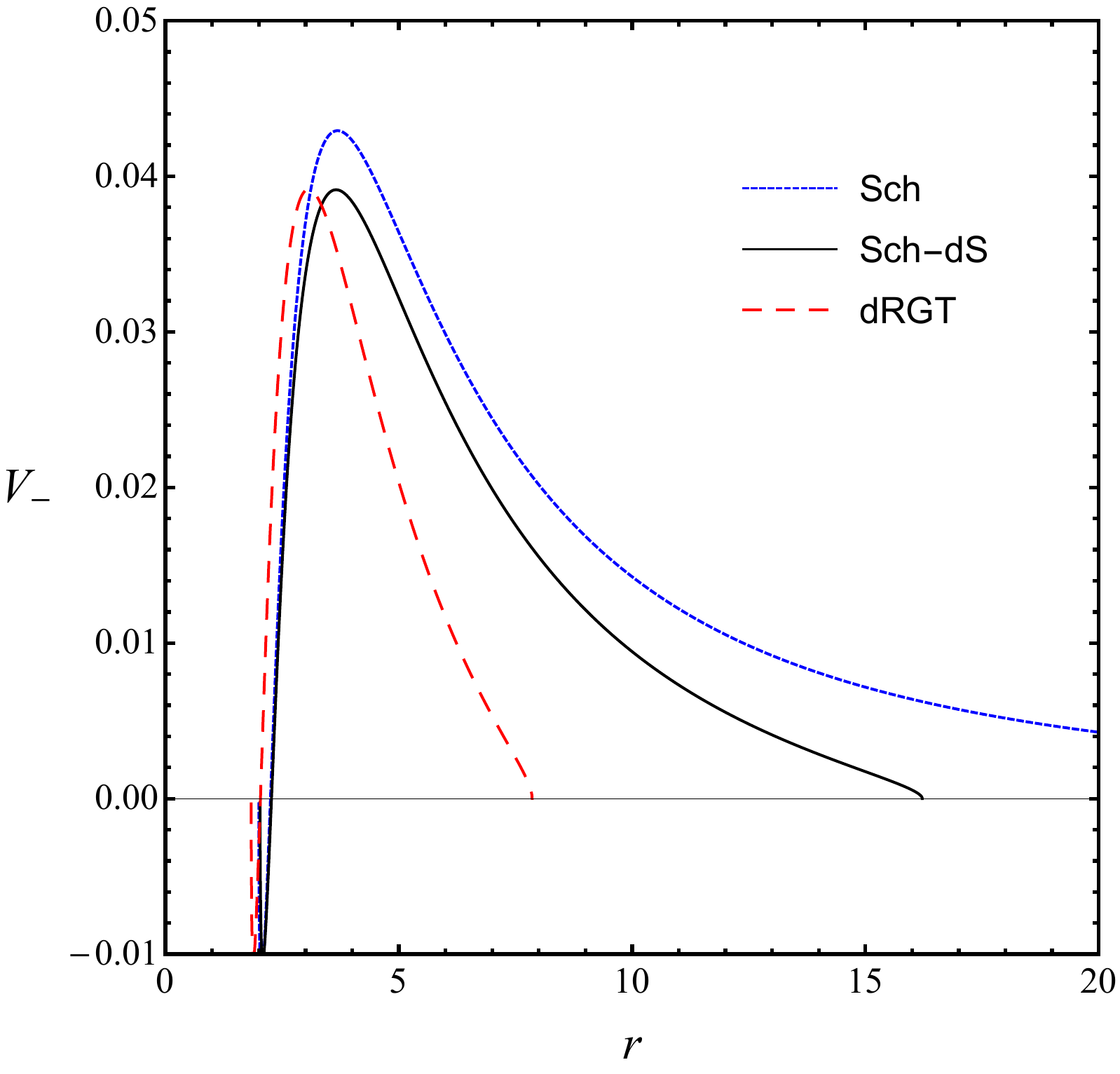}
\end{subfigure}
\begin{subfigure}{0.4\textwidth}
\includegraphics[width=\textwidth]{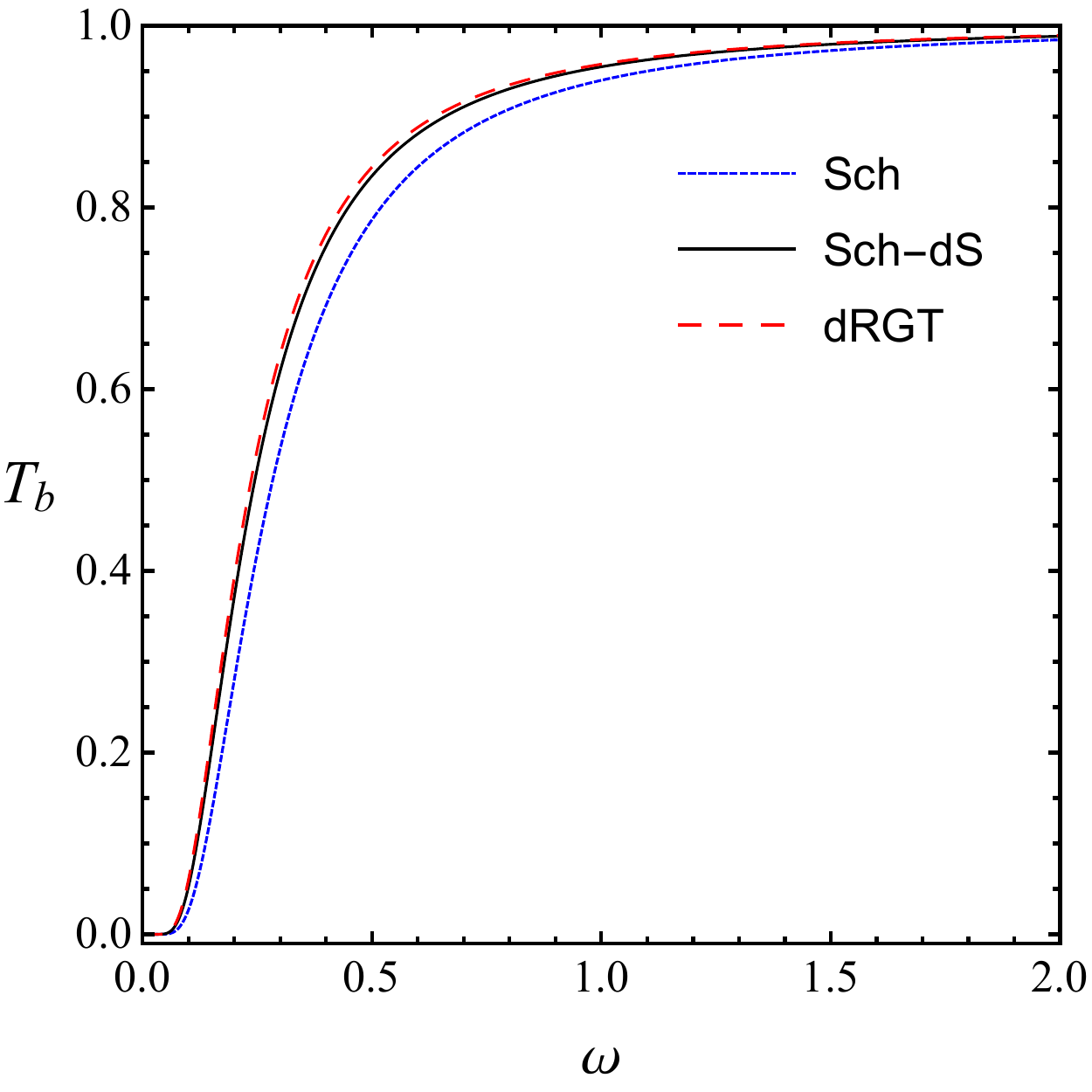}
\end{subfigure}
\end{center}
{\caption{The left panel shows the potential for Sch, Sch-dS and dRGT black holes with $\alpha_g = M=1, c_2 = -1/300, \beta_m = 0.755$ and $l=0$. The right panel shows the corresponding greybody factor bound.}\label{fig:VT-compare}}
\end{figure}

For the massless case, one can take $m=0$, then the integral in Eq. \eqref{Ab} can be written as
\begin{equation}
\int_{-\infty}^{\infty}\left|W^{2}\right|dr_{*} = \int_{r_{H}}^{R_{H}}\frac{\lambda^{2}}{r^{2}}dr = \lambda^{2}\left(\frac{1}{r_{H}} - \frac{1}{R_{H}}\right).
\end{equation}
By substituting the result of this integral in Eq. \eqref{TT}, the rigorous bound can be expressed as
\begin{equation}
T_b = \text{sech}^{2}\left(\frac{\lambda^{2}}{2\omega}\left[\frac{1}{r_{H}} - \frac{1}{R_{H}}\right]\right). \label{Tb-massless}
\end{equation}
One can see that the bound depends only on the distance between the horizons. This means that the result of the grey body factor bound depends only on the model parameters up to the event horizons. Then one can use this formulation for most kinds of black holes. Moreover, it is more general in the sense that it can also be applied to the scalar field case as seen in \cite{Boonserm:2017qcq}. Now let us compare the results for three kinds of black holes; Schwarzshild (Sch), Schwarzshild-de Sitter (Sch-dS) and dRGT black holes. For the Sch black hole, $R_H \rightarrow \infty$ so that the argument in function sech is lager than the others. Therefore, the greybody factor is lower than the others as shown in the right panel of Fig. \ref{fig:VT-compare}. For the dRGT massive gravity and the Sch-dS black holes, the existence of the graviton mass makes the horizons closer and also thinner by setting the height of the potentials as equal. As a result, the wave can transmit through the potential for the dRGT massive gravity easier than one for the Sch-dS so that the greybody factor is higher as shown in Fig. \ref{fig:VT-compare}.

\begin{figure}
\begin{center}
\begin{subfigure}{0.4\textwidth}
\includegraphics[width=\textwidth]{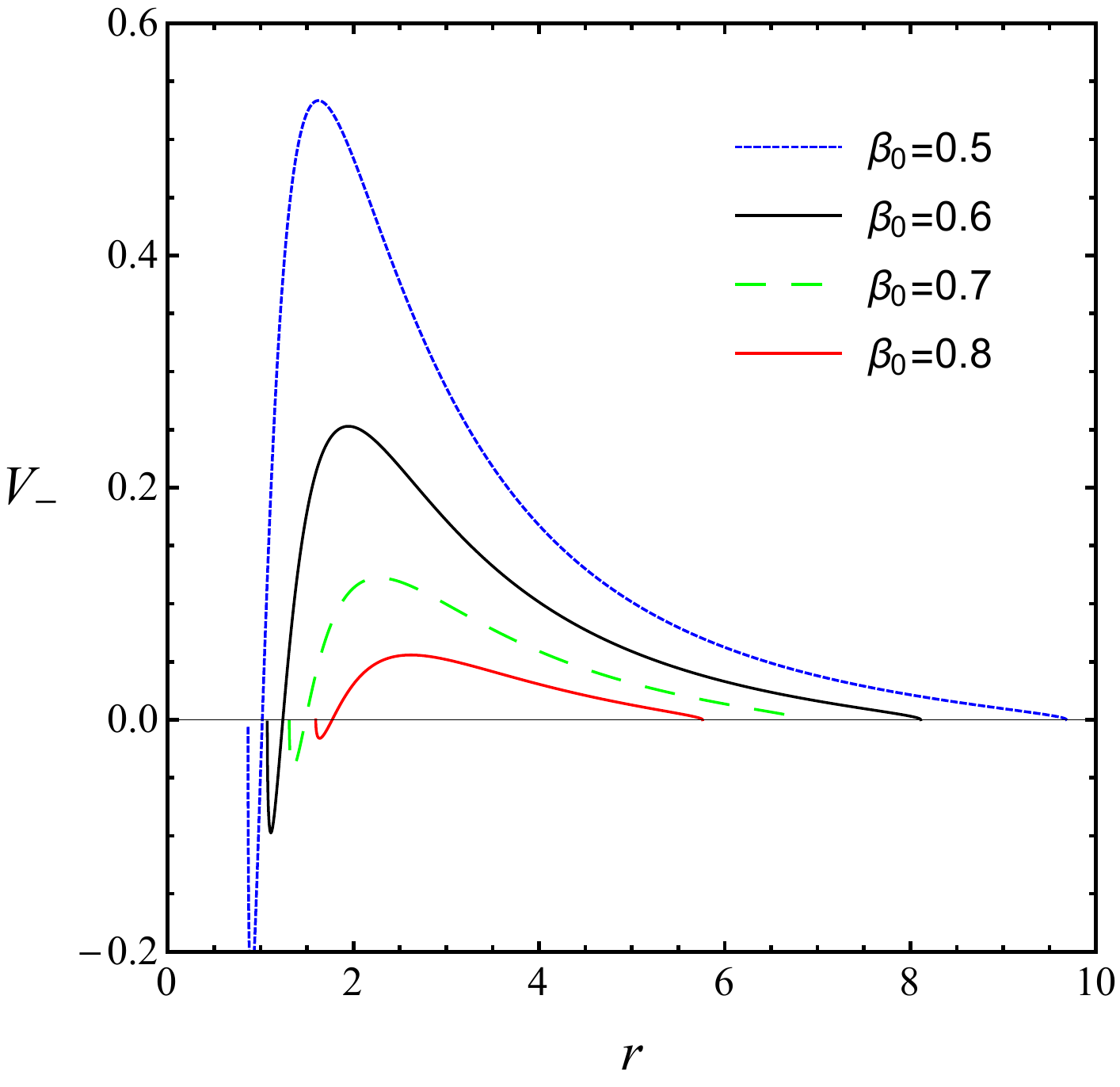}
\end{subfigure}
\begin{subfigure}{0.39\textwidth}
\includegraphics[width=\textwidth]{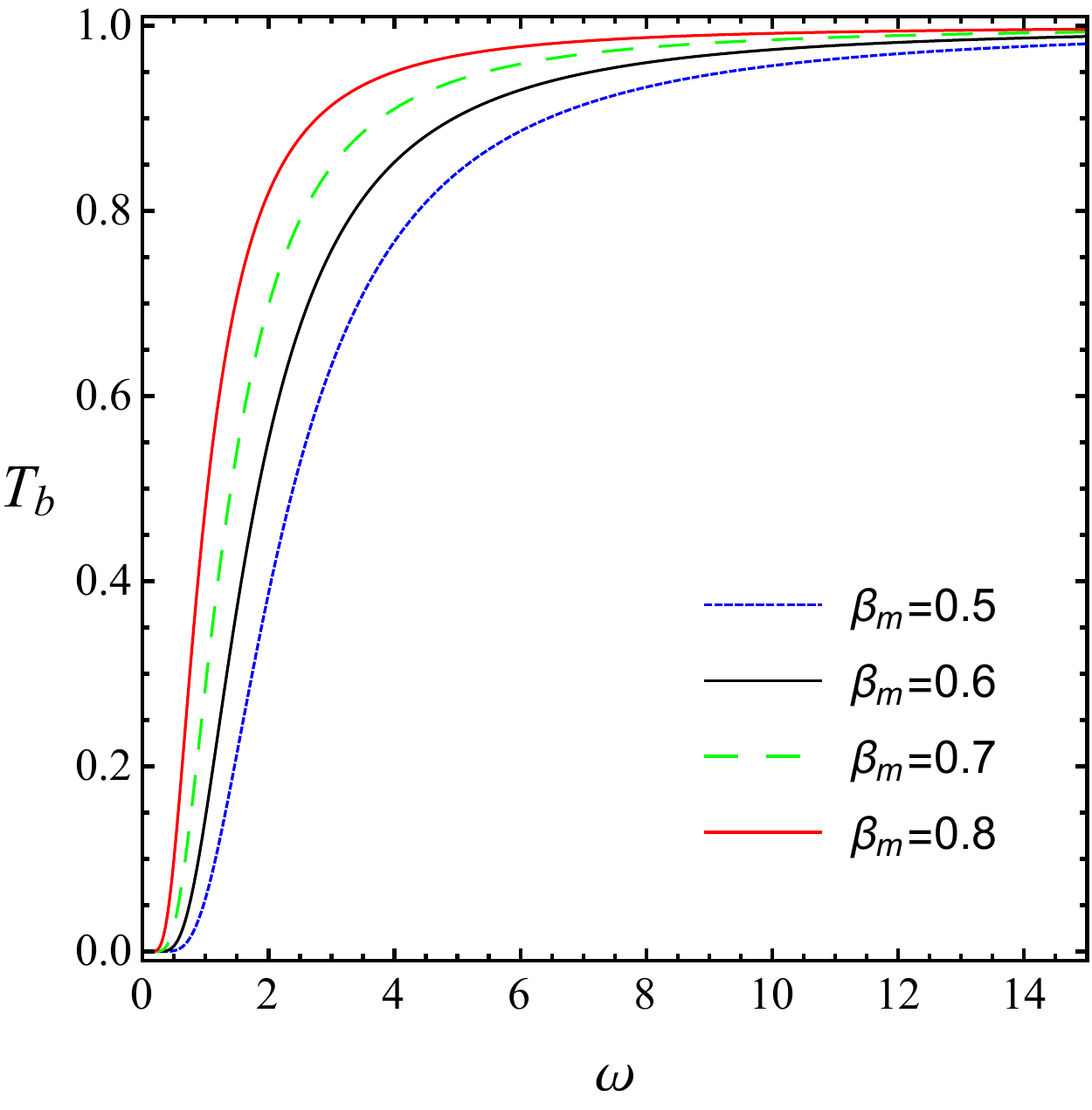}
\end{subfigure}
\end{center}
{\caption{The left panel shows the potential for dRGT black holes with $\alpha_g = M=1, c_2 = -2/300$ and $l=1$. The right panel shows the corresponding greybody factor bound.}\label{fig:VT-bm}}
\end{figure}

\begin{figure}
\begin{center}
\begin{subfigure}{0.4\textwidth}
\includegraphics[width=\textwidth]{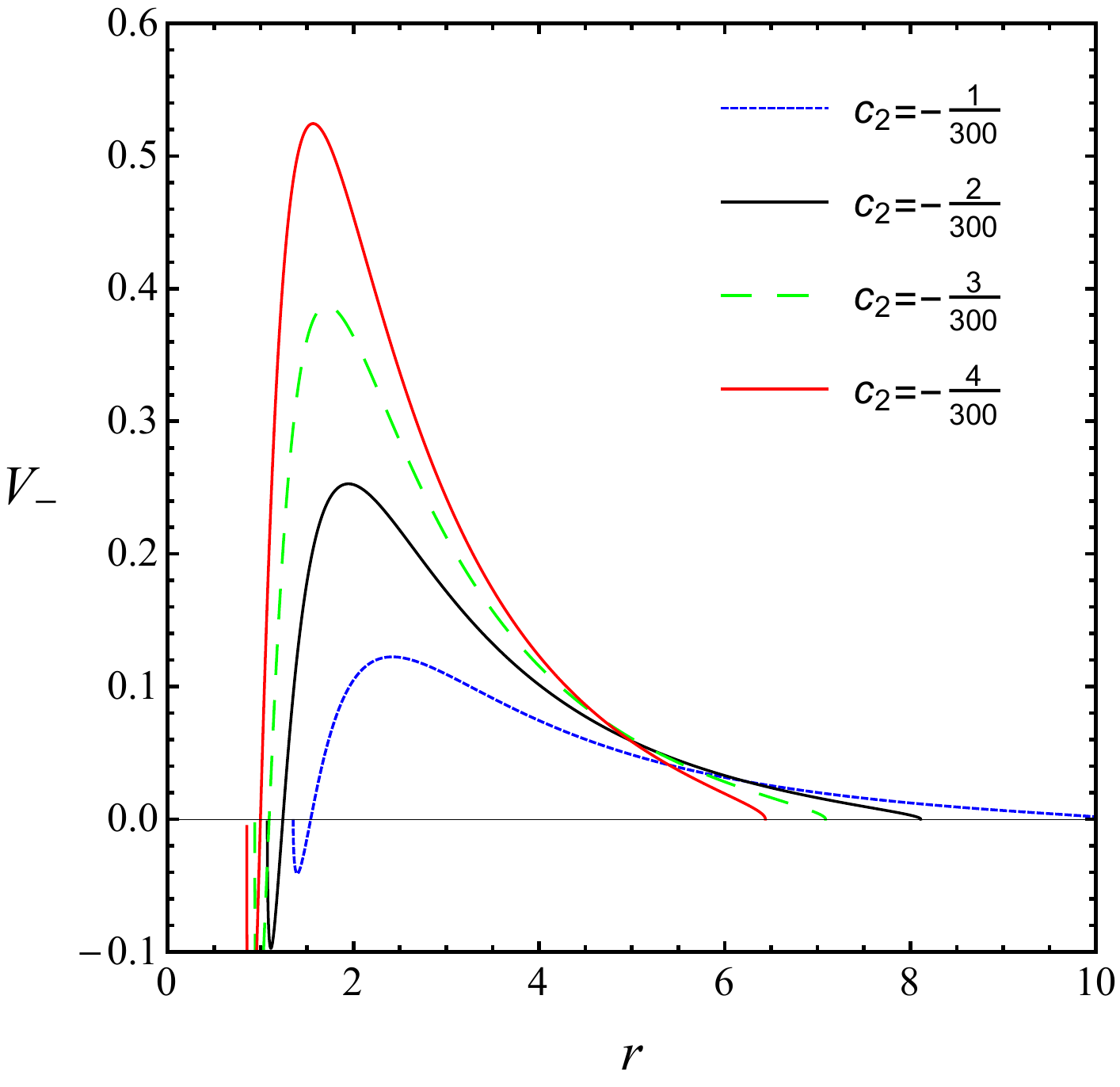}
\end{subfigure}
\begin{subfigure}{0.39\textwidth}
\includegraphics[width=\textwidth]{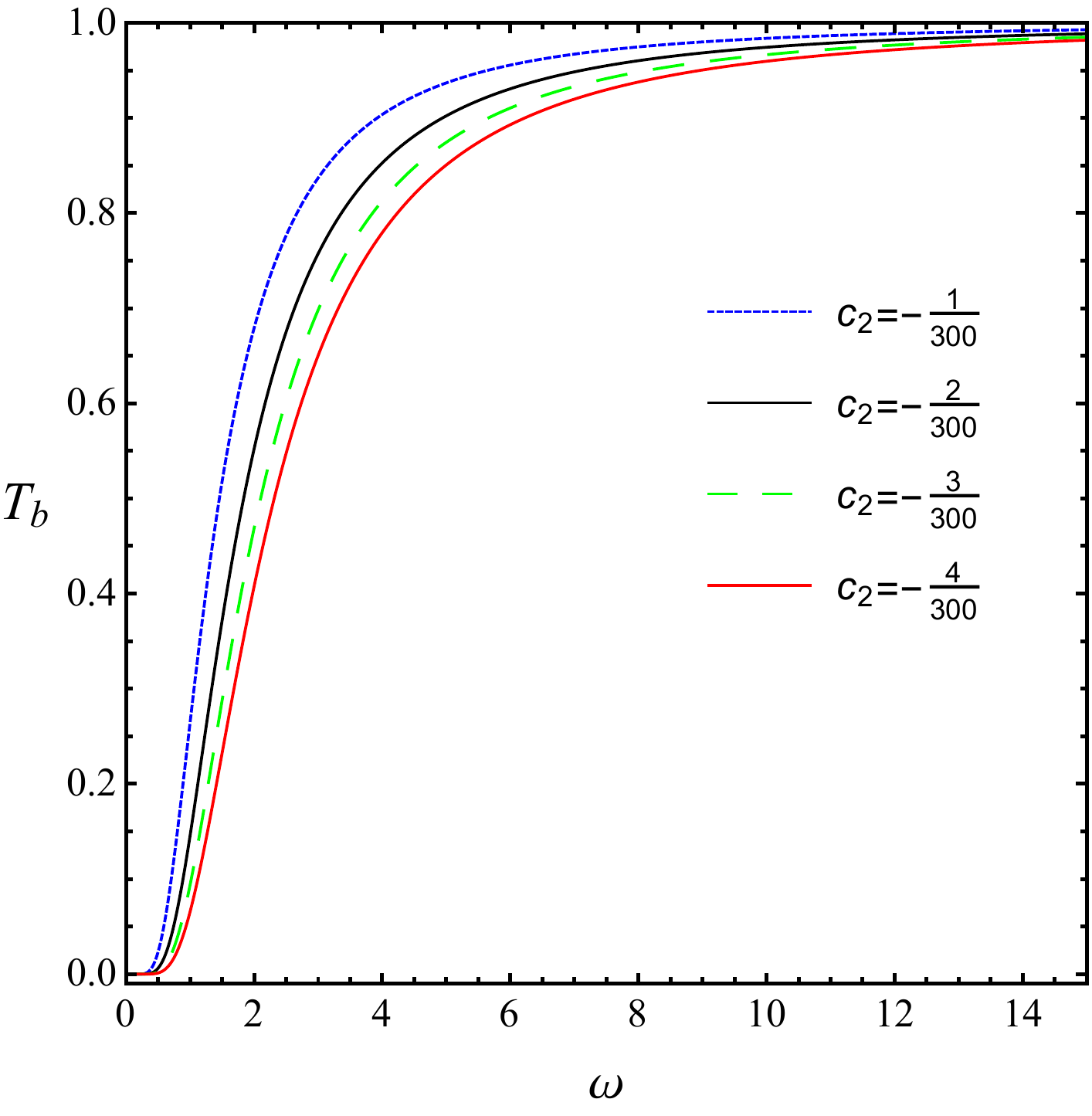}
\end{subfigure}
\end{center}
{\caption{The left panel shows the potential for dRGT black holes with $\alpha_g = M=1, \beta_m = 0.6$, and $l=1$. The right panel shows the corresponding greybody factor bound.}\label{fig:VT-c2}}
\end{figure}

For the behaviour of the greybody factor bound in dRGT massive gravity, one can use the same strategy since the argument of function sech depends on the distance between two horizons $\Delta r = R_H - r_H$. Therefore, the greybody factor bound will be large if $\Delta r$ is small. As a result, one can analyze the behaviour of the bound by using the fact that how $\Delta r$ depends on the parameters $c_2$ and $\beta_m$. These can be illustrated in Fig. \ref{fig:VT-bm} and Fig. \ref{fig:VT-c2}.

\begin{figure}
\begin{center}
\begin{subfigure}{0.4\textwidth}
\includegraphics[width=\textwidth]{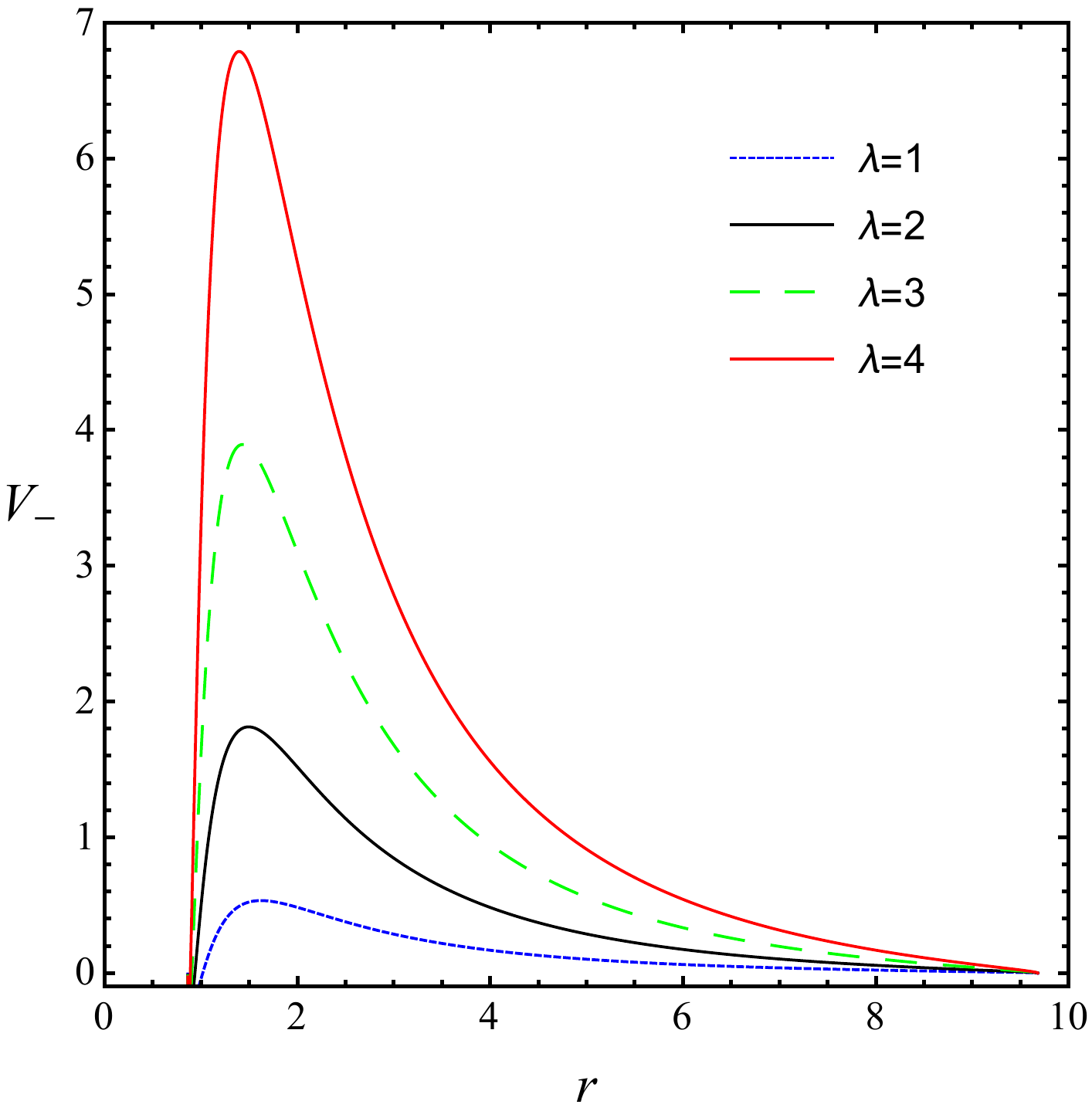}
\end{subfigure}
\begin{subfigure}{0.4\textwidth}
\includegraphics[width=\textwidth]{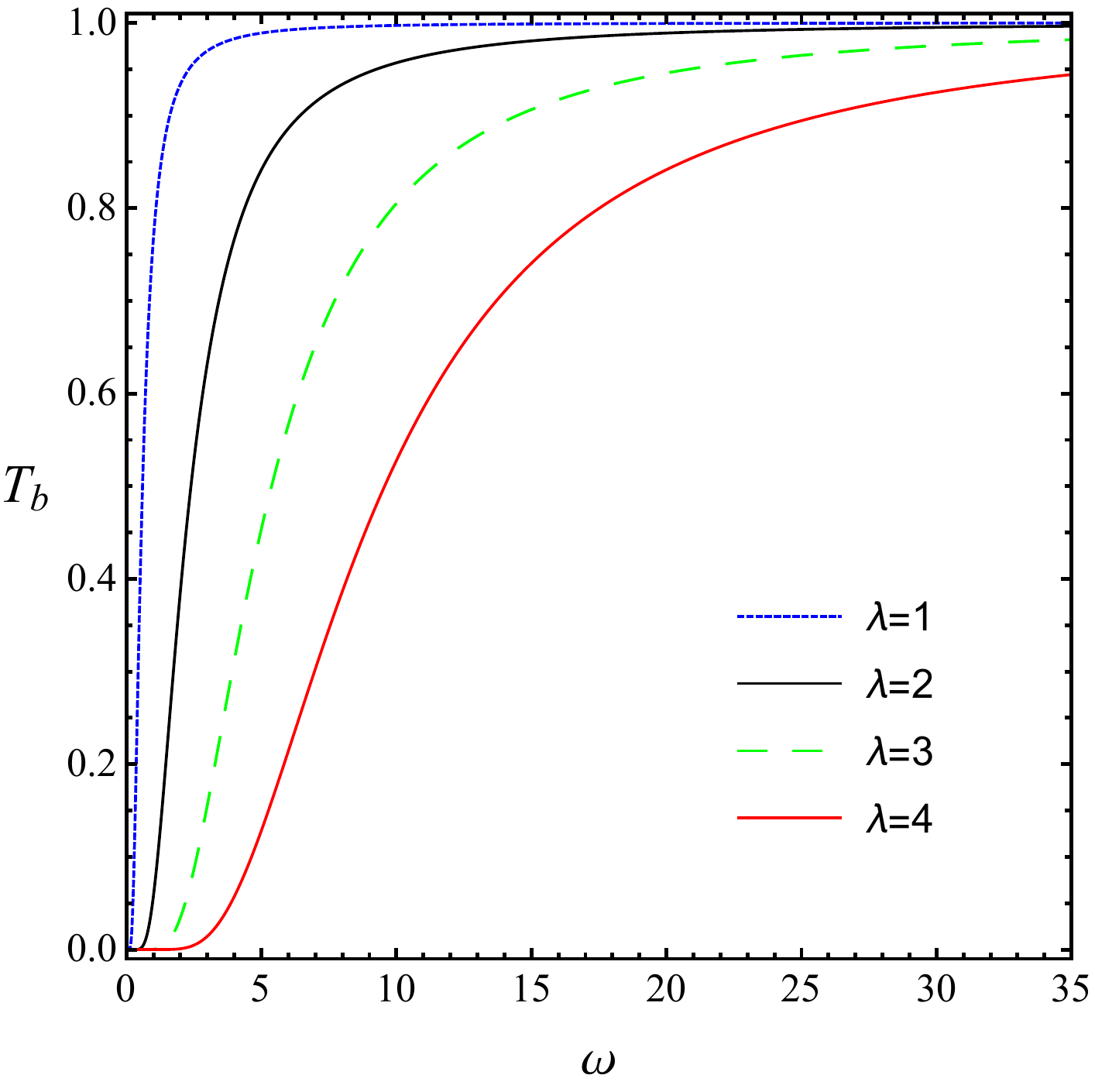}
\end{subfigure}
\end{center}
{\caption{The left panel shows the potential for dRGT black holes with $\alpha_g = M=1, c_2 = -2/300$ and $\beta_{m}=0.5$. The right panel shows the corresponding greybody factor bound.}\label{fig:VT-l}}
\end{figure}

Now we can analyze how the behaviour of $T_b$ depends on the shape of the potential. This can be done by varying the graviton mass parameters, $\beta_m$ and $c_2$, as well as the angular parameter $\lambda$. By fixing $c_2 = -2/300$ and $\beta_m = 0.5$, the potential gets higher when $\lambda$ increases as shown in the left panel in Fig.\ref{fig:VT-l}. The greybody factor becomes lower for a given value of $\omega$ since the wave is more difficult to transmit through the higher potential as shown in the right panel of Fig.\ref{fig:VT-l}. By fixing $c_2 = -2/300$ and $\lambda = 2$, the potential gets higher when $\beta_m$ decreases as shown in the left panel in Fig.\ref{fig:VT-bm}. As a result, the bound of the greybody factor becomes lower since it is more difficult for the wave to go through the higher potential as shown in the right panel of Fig. \ref{fig:VT-bm}. By fixing $\beta_m$ and $\lambda$, the analysis can be evaluated in the same way and can be seen in Fig. \ref{fig:VT-c2}.

One can see that the behaviour of the greybody factor can be analyzed as conducted in quantum mechanics, even though the physical situations are different. In quantum mechanics, the wave is supposed to pass the potential, while in this situation, the wave is supposed to occur near the black holes and then escape from the black hole. In this case, the ability of the wave to escape the black hole can be characterized from the curvature of the spacetime, which acts like the barrier to obstruct the wave.

\subsection{massive fermion}
\begin{figure}[h!]
\begin{center}
\includegraphics[scale=0.50]{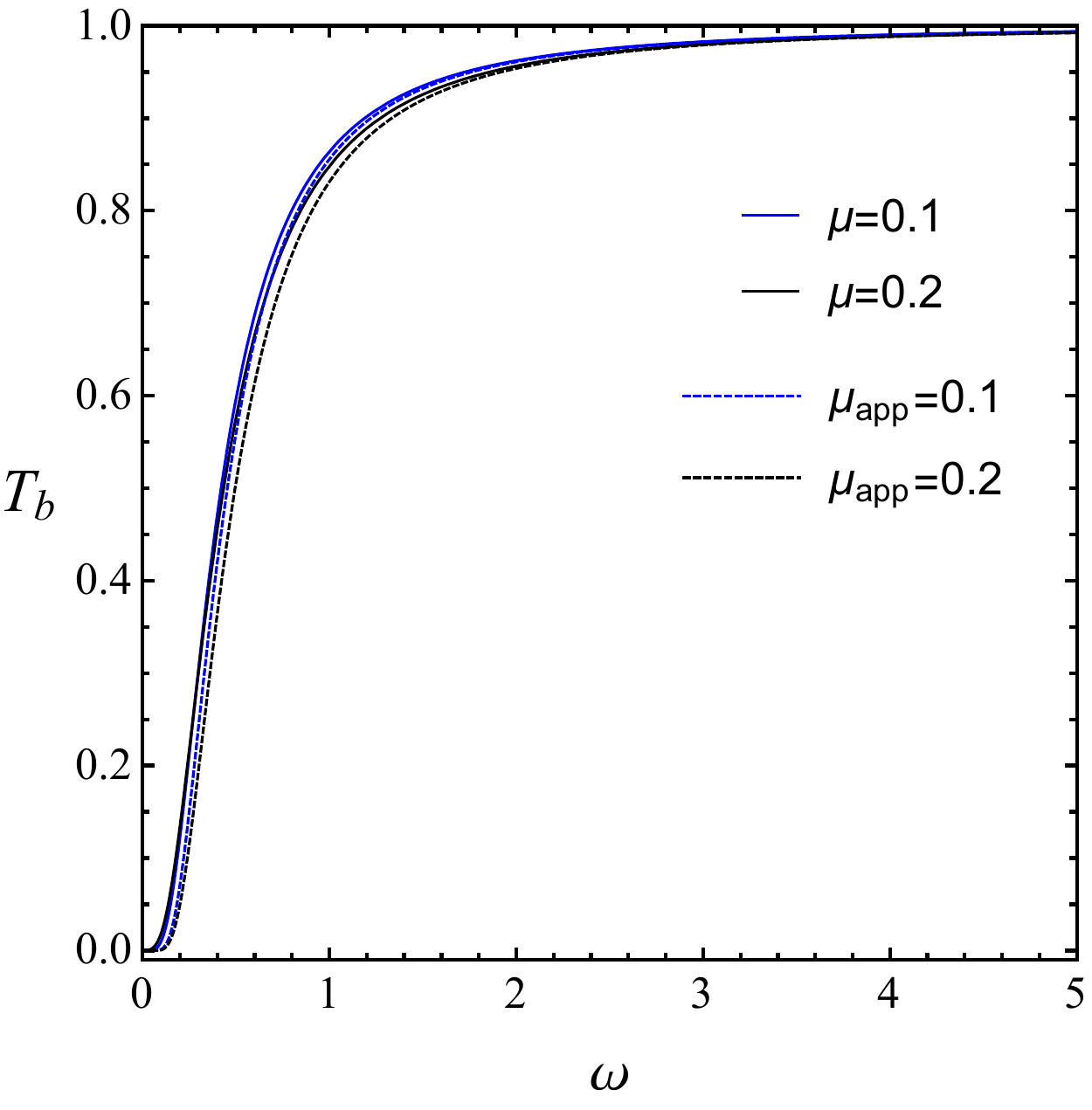}\qquad
\includegraphics[scale=0.50]{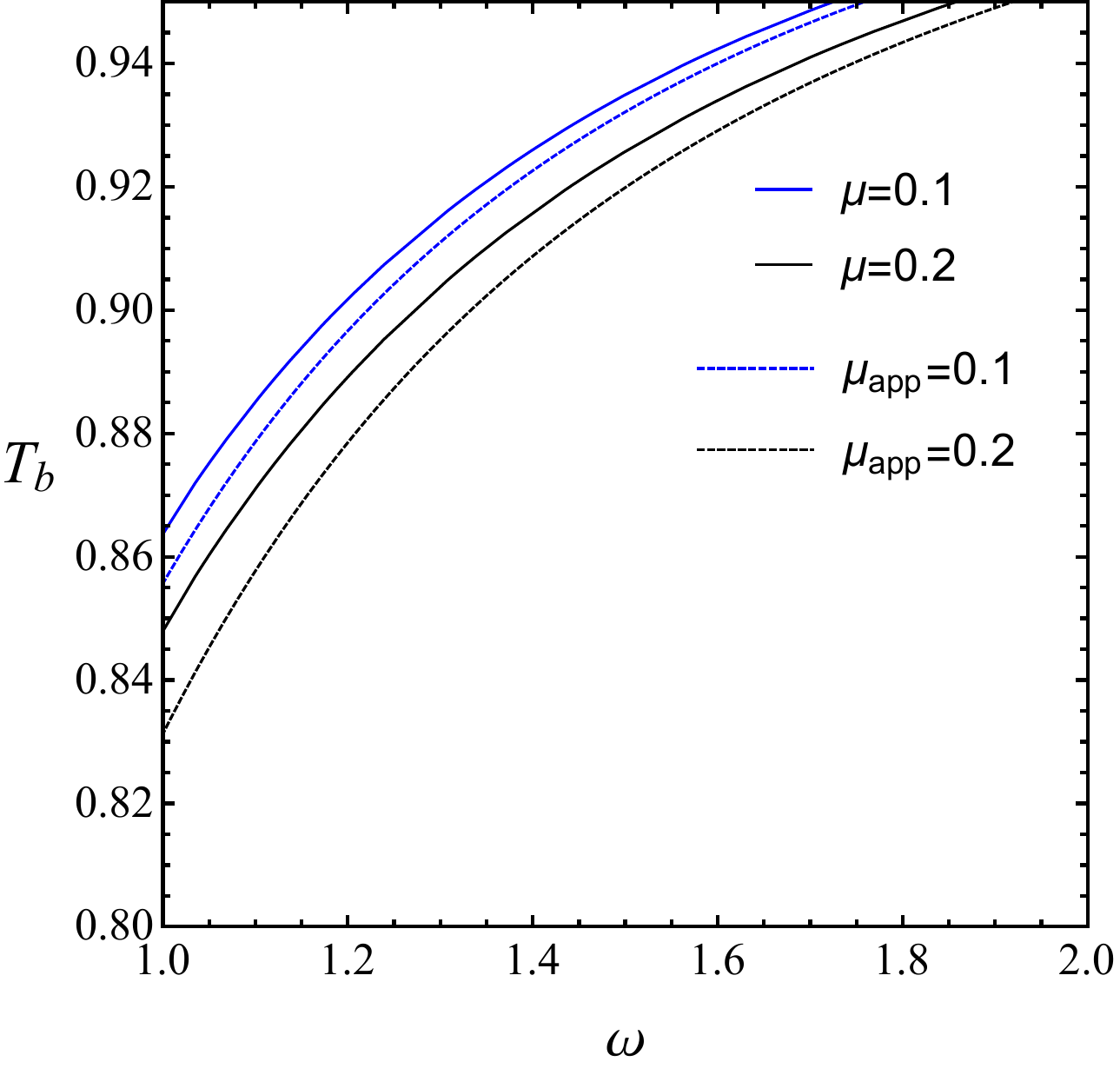}
\end{center}
{\caption{The greybody factor bound is evaluated using the full expression (solid line) and approximated expression (dashed line) with $\alpha_g = M=1, \beta_m = 0.8, c_2 = -0.1/3$.}\label{fig:Tappf}}
\end{figure}

For the massive case, one has to evaluate the full integral in Eq. \eqref{Ab}. The solution can be analytically obtained, however, it is complicated to analyze the behaviour of the greybody factor. We have shown the derivation and the integration results in Appendix. \ref{AppA}. In this section, we use the approximation in order to analyze the greybody factor bound. For the region between the two horizons, we have $f > 0$. Then the integral in Eq. \eqref{Ab} can be expressed as
\begin{eqnarray}
\int_{-\infty}^{\infty}\left|W^{2}\right|dr_{*} &=& \int_{r_{H}}^{R_{H}}\frac{1}{r^{2}}\frac{\left(\lambda^{2} + m^{2}r^{2}\right)^{2}}{\lambda^{2} + m^{2}r^{2} + \left(\lambda m/2\omega\right)f}dr\nonumber\\
     &=& \int_{r_{H}}^{R_{H}} \frac{\lambda ^2 \left(1+\mu ^2 r^2\right)}{r^2 \left(1+\frac{f \mu }{2 \omega  \left(\mu ^2 r^2+1\right)}\right)} dr = \int_{r_{H}}^{R_{H}}A dr,
\end{eqnarray}
where
\begin{eqnarray}
          A &=& \frac{\lambda ^2 \left(1+\mu ^2 r^2\right)}{r^2 \left(1+\frac{f \mu }{2 \omega  \left(\mu ^2 r^2+1\right)}\right)},\quad
          \mu =\frac{m}{\lambda}.
\end{eqnarray}
Considering the integrand $A$, one can see that the factor $\left(1+\frac{f \mu }{2 \omega  \left(\mu ^2 r^2+1\right)}\right)$ is larger than $1$. Therefore, one can use this inequality to approximate the new integrand, which corresponds to the new greybody factor bound. As a result, the integrand can be written as follows
\begin{eqnarray}
A &=& \frac{\lambda^2}{r^2}\frac{ \left(1+\mu ^2 r^2\right)}{ \left(1+\frac{f \mu }{2 \omega  \left(\mu ^2 r^2+1\right)}\right)} \leq \frac{\lambda^2}{r^2}\left(1+\mu ^2 r^2\right) =A_{app}.
\end{eqnarray}
Since $A$ and $A_{app}$ are positive functions for the range $r_H < r < R_H$, it is found that the integral can be expressed as $\int A dr \leq \int A_{app} dr$. As a result, one obtains the greybody bound as follows
\begin{eqnarray}
T \geq \text{sech}^{2}\left( \int_{r_{H}}^{R_{H}}A dr \right) \geq  \text{sech}^2\left( \int_{r_{H}}^{R_{H}}A_{app} dr \right) =T_b.
\end{eqnarray}
This behaviour can be seen explicitly in Fig. \ref{fig:Tappf}. From this figure, one can see that it is possible to take $A_{app}$ to evaluate the greybody factor bound. Moreover, it provides us with a useful way to analytically discuss the behaviour of the greybody factor,
\begin{eqnarray}
T_b &=&  \text{sech}^2\left( \int_{r_{H}}^{R_{H}}A_{app} dr \right), \\
&=& \text{sech}^{2}\left(\frac{\lambda^{2}}{2\omega}\left[\frac{1}{r_{H}} - \frac{1}{R_{H}}+\mu^2 (R_H -r_H)\right]\right),\\
&=& \text{sech}^{2}\left(\frac{\lambda^{2}}{2\omega}\frac{(R_H - r_H)}{R_H r_H}\left[1+\mu^2 R_H r_H\right]\right).\label{Tapp1}
\end{eqnarray}

From this expression, one can see that the greybody factor bound reduces to one for the massless case where $\mu \rightarrow 0$. Moreover, it is very useful since the argument of the function sech is still proportional to $R_H -r_H$. This implies that the bound for the massive case is still dependent on the model parameters $c_2$ and $\beta_m$ in the same way as one for the massless case. Particularly, the greybody factor bound will be large if the distance between two horizons is small. Likewise, it will be large if the magnitude of $c_2$ is small or $\beta_m$ is close to $1$. This behaviour can also be found when using the full expression. Since this behaviour does not significantly differ from the massless case, we omit to show the numerical plots explicitly for brevity. It is worthwhile to note that even though the approximation is valid for the entire range of parameter $\mu$, our evaluation is performed by keeping $\mu$ small. It is sufficient to avoid the backreaction for using limit.

\begin{figure}[h!]
\begin{center}
\includegraphics[scale=0.50]{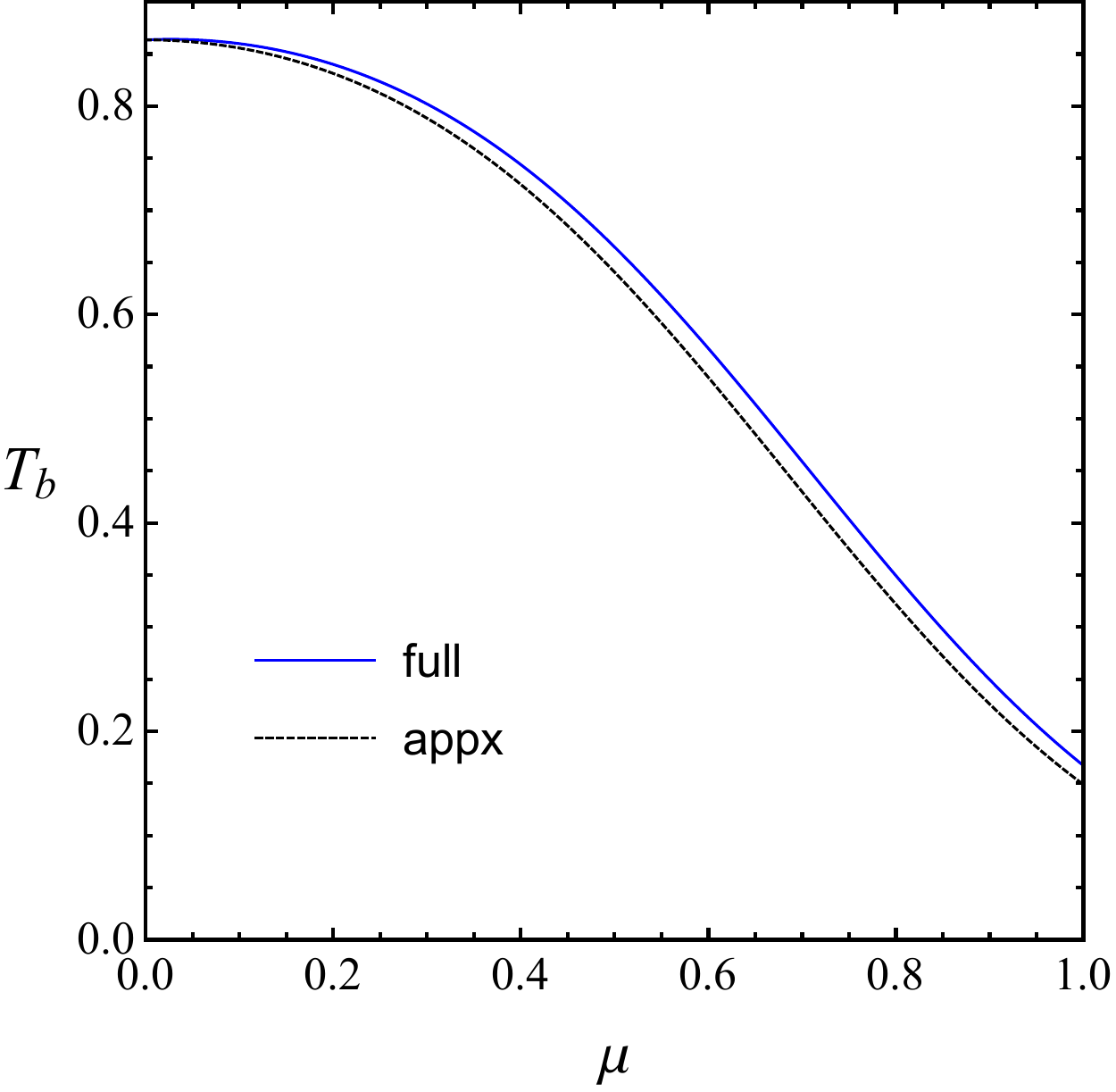}\qquad
\includegraphics[scale=0.52]{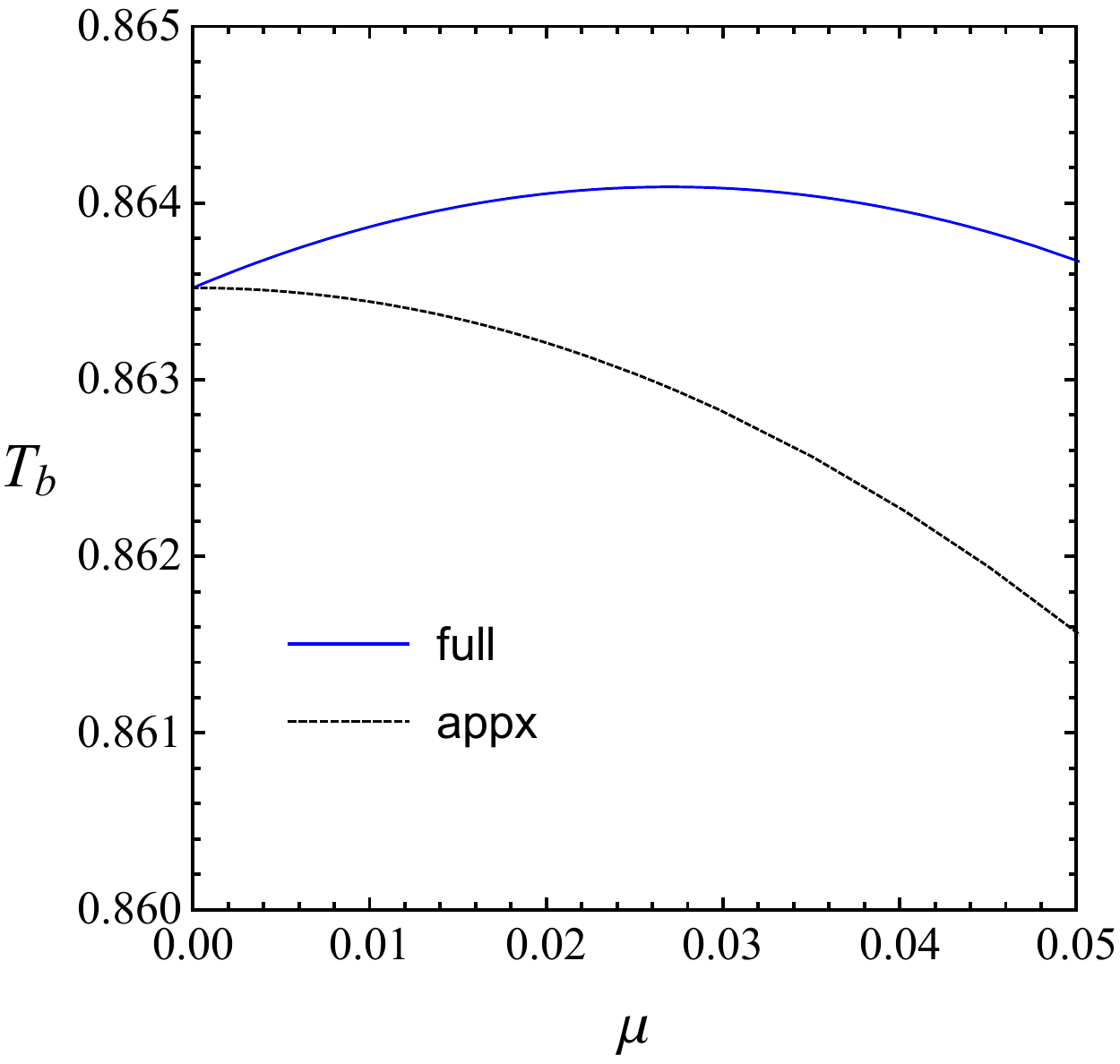}
\end{center}
{\caption{The greybody factor bound is evaluated using the full expression (solid-blue line) and approximated expression (dashed-black line) with $\alpha_g = M=1, \beta_m = 0.8, c_2 = -0.1/3, \omega=1.$}\label{fig:Tmufullapp}}
\end{figure}

From Eq. \eqref{Tapp1}, it is found that the greybody factor bound for the massive case seems to be less than one for the massless case. In other words, the higher the value of $\mu$ the lower the bound of the greybody factor as seen in the left panel on Fig. \ref{fig:Tmufullapp}. This argument points out that the massive particles will have more self interactions, which will then make it more difficult for the particles to pass through the potential. However, this is not valid for very small values of $\mu$ as shown in the right panel of Fig. \ref{fig:Tmufullapp}. From this figure, we found that the new bound is still valid, but some information is lost. There exists an extremal point for the full expression, but not for the approximated expression. Actually, the argument of sech function is proportional to $\mu^2$ so that there are no extrema.

In order to improve the bound by taking into account such an effect, one may add more terms into the approximated expression. This may be performed by using the series expansion of  the factor $\left(1+\frac{f \mu }{2 \omega  \left(\mu ^2 r^2+1\right)}\right)$ with keeping $\mu$ small. As a result, the improved bound can be obtained as follows
\begin{eqnarray}
A &\approx & \frac{\lambda^2}{r^2}\left(1 -\frac{f}{2\omega}\mu+ \left(r^2 + \frac{f^2}{4\omega^2}\right)\mu^2 -\frac{f^3\mu ^3  }{8\omega ^3}\right),\nonumber \\
&\lesssim &  \frac{\lambda^2}{r^2}\left(1 -\frac{f}{2\omega}\mu+ \left(r^2 + \frac{f^2}{4\omega^2}\right)\mu^2 \right)\equiv A_{app}. \label{Aapp2}
\end{eqnarray}

\begin{figure}[h!]
\begin{center}
\includegraphics[scale=0.50]{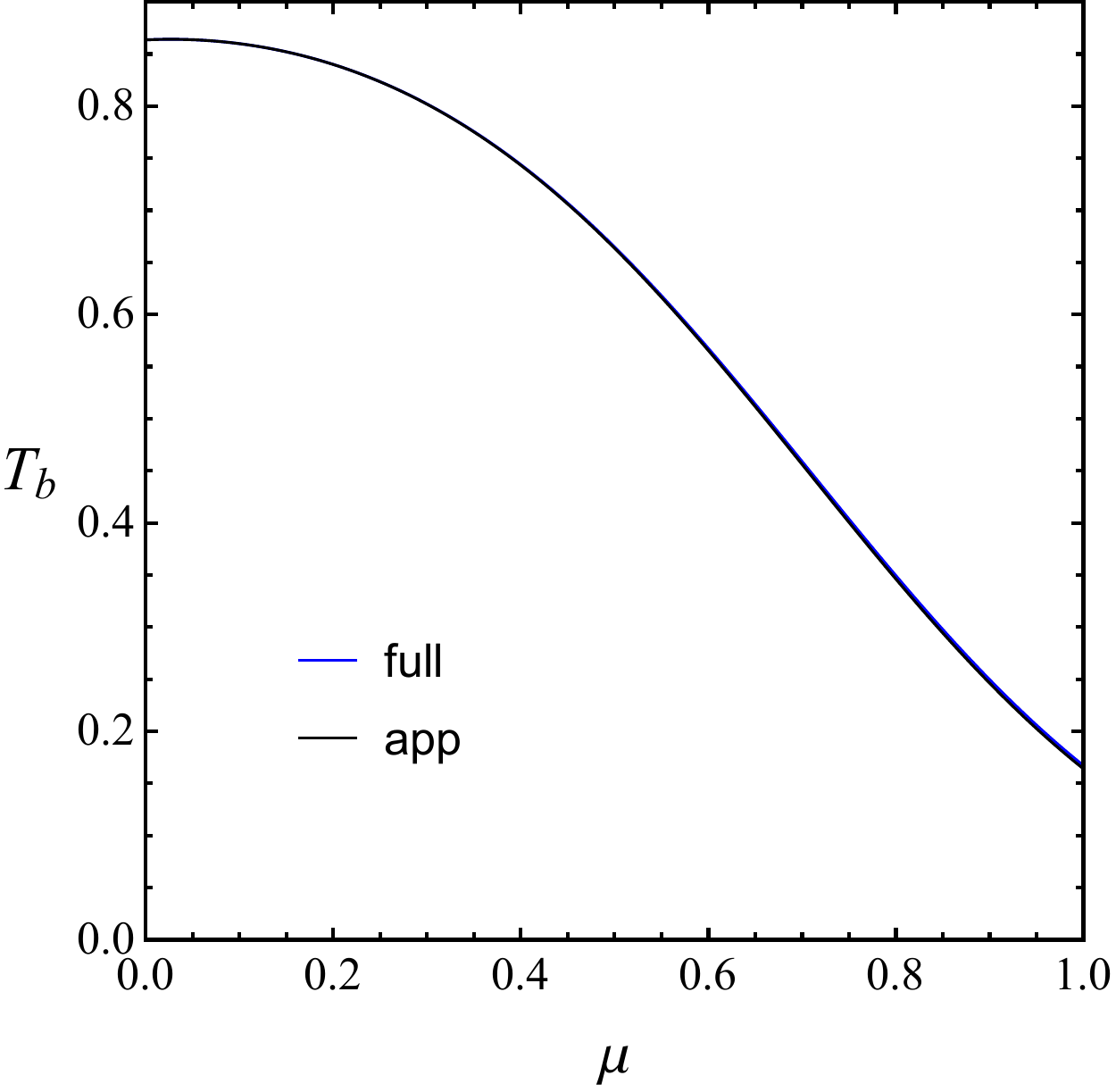}\qquad
\includegraphics[scale=0.50]{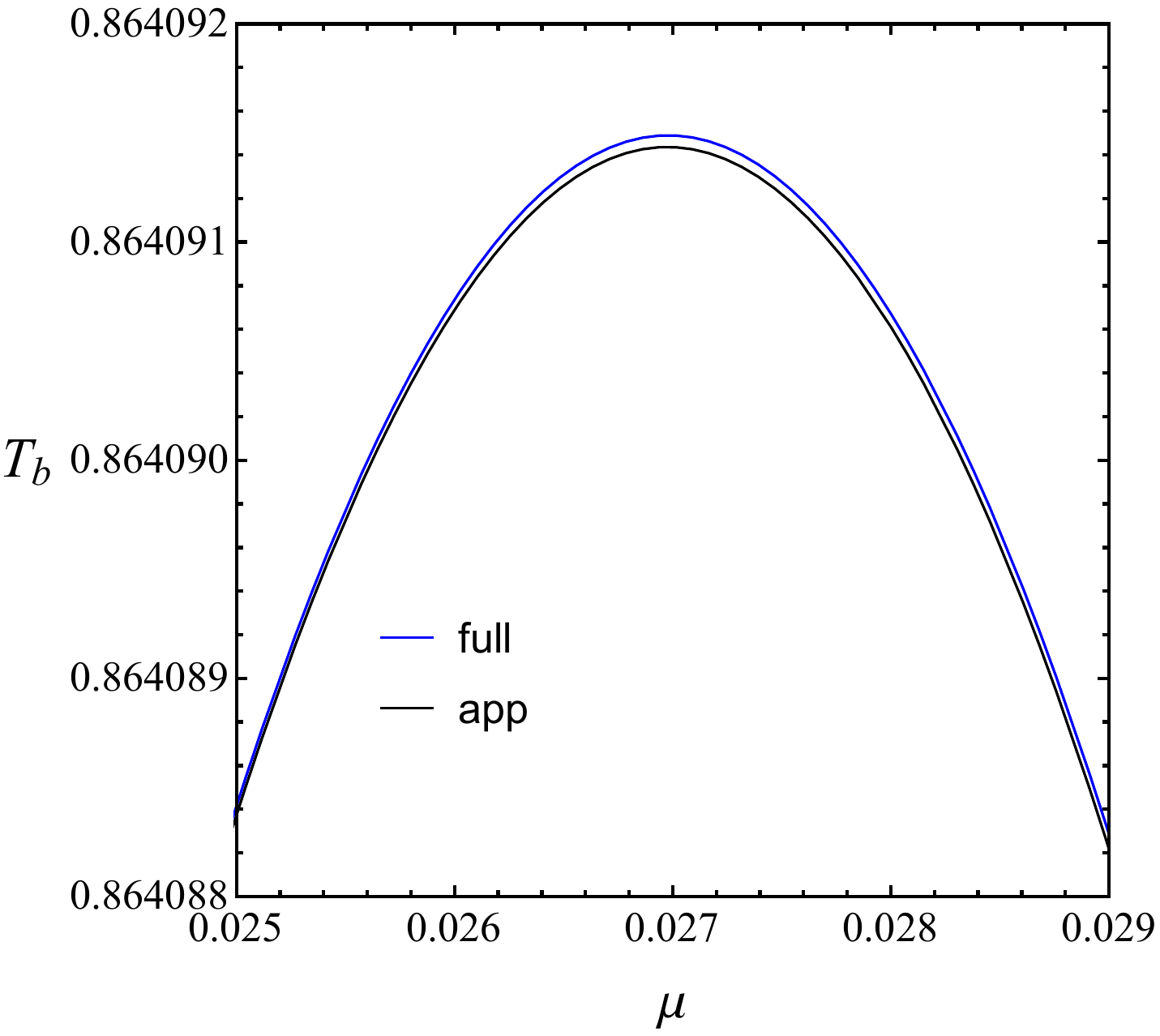}
\end{center}
{\caption{The greybody factor bound is evaluated using the full expression (solid-blue line) and approximated expression (dashed-black line) with $\alpha_g = M=1, \beta_m = 0.8, c_2 = -0.1/3, \omega=1.$}\label{fig:Tmufullapp2}}
\end{figure}
\begin{figure}[h!]
\begin{center}
\includegraphics[scale=0.450]{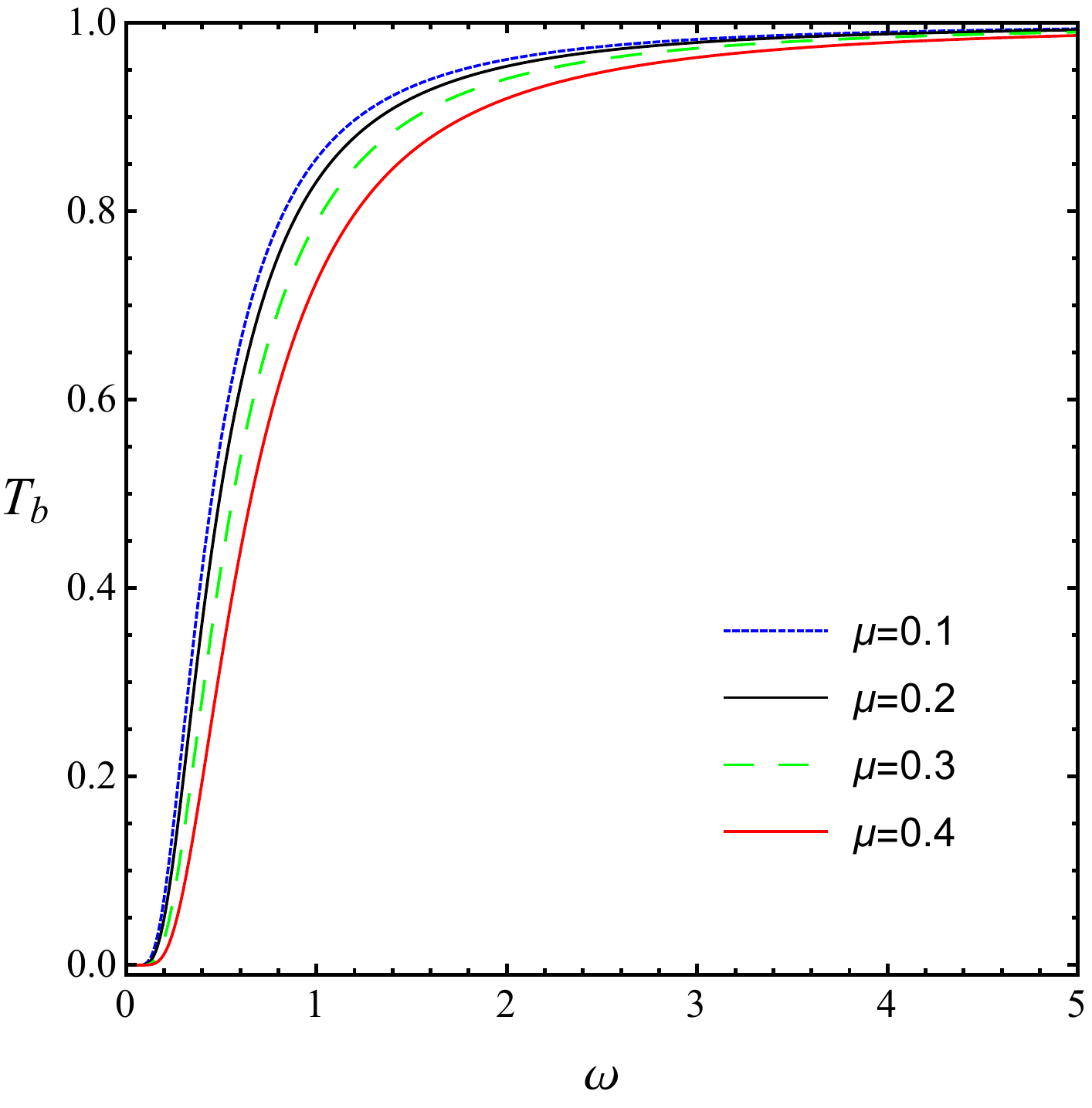}\qquad
\includegraphics[scale=0.470]{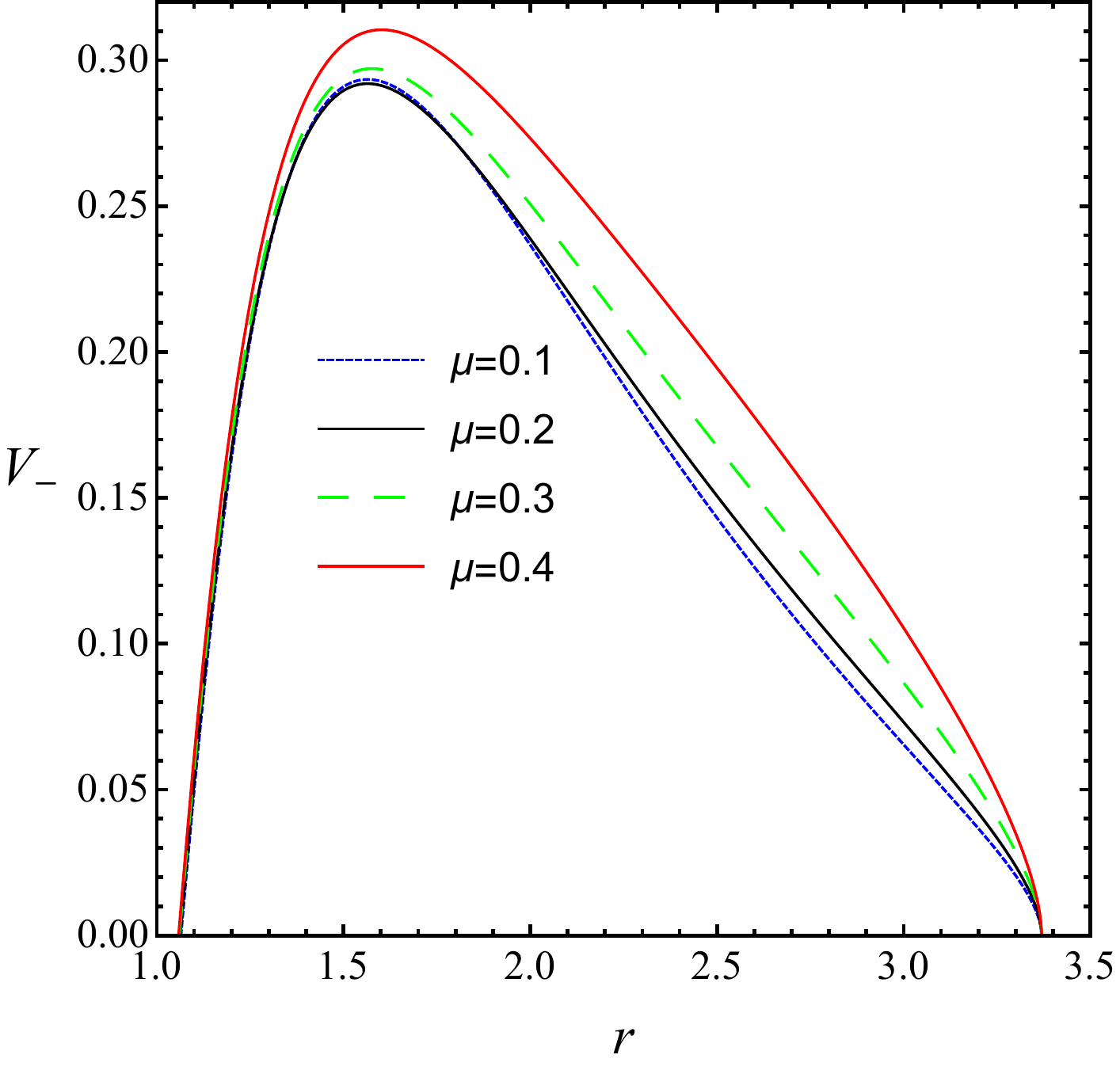}
\end{center}
{\caption{The greybody factor bound (left panel) and  the potential with various value of $\mu=m/\lambda$ where $\alpha_g = M=1, \beta_m = 0.8, c_2 = -0.1/3$.}\label{fig:TVm-mu}}
\end{figure}

The second line from the above equation can be obtained by the fact that the third order is always negative. One can see that the integrand can be integrated easily since it is just a polynomial function of $r$. Substituting this expression into the definition of the greybody factor bound, one obtains
\begin{eqnarray}
T_b &=& \text{sech}^{2}\left(\frac{\lambda^{2}}{2\omega}\left[\frac{(R_H - r_H)}{R_H r_H}-F_1 \mu+ F_2 \mu^2\right]\right),\label{Tapp2}
\end{eqnarray}
where the functions $F_1$ and $F_2$ are the resulting integration as
\begin{eqnarray}
F_1 &=& \int_{r_{H}}^{R_{H}}\frac{f}{2\omega r^2}dr,\quad
F_2 = \int_{r_{H}}^{R_{H}}\left(1+\frac{f^2}{2\omega r^2} \right)dr.
\end{eqnarray}
By performing numerical investigation, we found that the greybody factor bound from the approximated expression in Eq. \eqref{Tapp2} is very close to the full expression as shown in Fig. \ref{fig:Tmufullapp2}. Moreover, it provides the critical point $\mu_c$ at nearly the same point with the full expression at
\begin{eqnarray}
\mu_c = \frac{F_1}{2 F_2}. \label{mut}
\end{eqnarray}
This point can be obtained by maximizing the bound in Eq. \eqref{Tapp2}. For example, by using the same parameter setting as one in Fig. \ref{fig:Tmufullapp2}, we have $\mu_c = 0.0269753$. In Fig. \ref{fig:TVm-mu}, we show how the behaviour of the greybody factor bound depends on the shape of the potential. It is found that the greybody factor will be large when the potential is small. Note that peak of the potential for $\mu = 0.1$ is higher than one for $\mu = 0.2$. However, the potential for $\mu = 0.1$ is thinner than one for  $\mu = 0.2$ so that there is a greater probability for the particle to pass through the potential, therefore, the greybody factor is higher. Note also that there exists a critical point for the potential as found in the right panel of Fig. \ref{fig:TVm-mu}, but it is not the same critical point for the greybody factor as we have discussed. This may be one of the disadvantages of the rigorous bound method. The bound is still valid, but some tiny effect may be lost. Moreover, for large $\lambda$, the bound is much lower than the exact value obtained in other methods. Then this method may not be useful for the large multipole $\lambda$. We will address this issue in the next section by comparing the result to one when using the WKB method.

\section{Greybody factors using the WKB approach} \label{WKB}

In this section, we investigate the greybody factor based on the third order WKB approximation proposed by Iyer and Will \cite{Iyer:1986np}. This is a well-developed method to study the barrier-like Quasi-normal modes and the greybody factor. From Sec. (\ref{eom}), it is found that the crucial property of the potential is that the potential depends on the energy and mass of the Dirac field. This implies that the shape of the effective potential relates to the energy and mass of the considered particles. For this type of effective potentials, we need to consider series expansions on every step of evaluating the WKB approximation. This method was first studied by Simone and Will for the Quasi-normal modes of massive scalar perturbations in Schwarzschild and Kerr black holes \cite{Simone:1991wn}. A further generalization to massive Dirac perturbations in both the context of Quasi-normal modes and greybody factor for Schwarzschild spacetime were studied by Cho and Lin \cite{Cho:2003qe,Cho:2004wj}. It is worth to note that the recent update of the WKB methods for studying the greybody factor is improved to a higher order approximation by Konoplya, Zhidenko and Zinhailo \cite{Konoplya:2019hlu}. However, in our consideration, for the comparison of massless and massive Dirac field cases, it is more convenient for us to work in the third order approximation.


\subsection{The Methods}

The key idea of the WKB approximation is the use of series expansion to obtain the solution. The method is, therefore, crucially dependent on the shape of the potential as well as the energy of the particles being considered. As a result, in order to obtain the results analytically, one can separate the specific considerations case by case; for example, $\omega^2 \sim V$ or $\omega^2 \ll V$. In this subsection, we will follow the analytical expression investigated in \cite{Cho:2004wj} and then separate the resulting greybody factor into two categories: Intermediate energy approximation and Low energy approximation.

\subsubsection{Intermediate energy approximation}\label{IEA}
For the WKB approximation, it is convenient to rewrite the redial equation in Eq.~(\ref{RE}) in a suitable form as
\begin{equation}
\left(\frac{d^{2}}{dr_{*}^{2}}+Q\right)\Psi=0,
\end{equation}
where $Q=\omega^{2}-V(\omega,m,r)$ and $V(\omega,m,r)$ are chosen as $V_{-}$  in Eq.~(\ref{veff}).
For the maximum value of the effective potential $V_{max}$, the condition for intermediate energy can be written as  $\omega^{2}\approx V_{max}(\omega,m)$, where $r_{max}$ represents the corresponding radial value at the maximum point of the effective potential. The greybody factor for this approximation is given by \cite{Cho:2004wj}
\begin{equation}\label{absor}
T=\frac{1}{1+\exp^{2S\left(\omega\right)}}.
\end{equation}
The function $S(\omega)$ can be expressed as
\begin{equation}\label{ss}
\begin{aligned}
S(\omega)  = & \pi k^{1/2}\left[\frac{1}{2}z_{0}^{2} + \left(\frac{15}{64}b^{2}_{3} - \frac{3}{16}b_{4} \right)z_{0}^{4} \right]\\
& + \pi k ^{1/2}\left[\frac{1155}{2048}b_{3}^{4} - \frac{315}{256}b_{3}^{2}b_{4} + \frac{35}{128}b^{2}_{4} + \frac{35}{64}b_{3}b_{5} - \frac{5}{32}b_{6} \right]z_{0}^{6} + \pi k^{-1/2}\left[\frac{3}{16}b_{4} -\frac{7}{64}b_{3}^{2} \right]\\
& - \pi k^{-1/2}\left[\frac{1365}{2048}b_{3}^{4} - \frac{525}{256}b_{3}^{2}b_{4} + \frac{85}{128}b_{4}^{2} + \frac{95}{64}b_{3}b_{5} - \frac{25}{32}b_{6} \right]z_{0}^{2}+O\left(\omega\right)\: .
\end{aligned}
\end{equation}
where $O\left(\omega\right)$ is the higher order terms and
\begin{equation}\label{sscoes}
\begin{aligned}
z_{0}^{2} = -\frac{Q_{max}}{k} \: ;\: \: k = \frac{1}{2}\left(\frac{d^{2}Q}{dr_{*}^{2}} \right)_{max} \: ;\: \:
  b_{n} = \left(\frac{1}{n!k} \right)\left(\frac{d^{n}Q}{dr_{*}^{n}} \right)_{max} \: .
\end{aligned}
\end{equation}
Note that the subscript ``$max$" denotes the quantities for $r=r_{max}$ after taking the derivative. Since the effective potential depends on the energy and mass of the Dirac field, we cannot obtain an exact value of $r_{max}$ where the value of multipole corresponding to the angular eigenvalue $\lambda$ is given. However, it is observed that the effective potential is independent of energy in the massless limit. Therefore, one can expand the potential around one for the massless case by keeping the fermion mass small. By adopting the new mass parameter as $\mu=m/\lambda$, the effective potential can be written as a series expansion:
\begin{equation}\label{vmu}
V\left(\omega,\mu,r\right)=V_{0}\left(r\right)+V_{1}\left(\omega,r\right)\mu+V_{2}\left(\omega,r\right)\mu^{2}+...+V_{n}\left(\omega,r\right)\mu^{n}.
\end{equation}
By using the same strategy, $r_{max}$ can be expanded as
\begin{equation}\label{rmu}
r_{max}=r_{0}+r_{1}\mu +r_{2}\mu^{2}+...+r_{n}\mu^{n}\equiv r_{0}+\delta,
\end{equation}
where $r_{0}$ denotes $r_{max}$ for the massless case, which is independent of $\omega$ and $\mu$. In order to obtain $r_{max}$, one needs to solve the following equations,
\begin{eqnarray}\label{rmaxcondi}
0&=&\partial_{r}V\left(\omega,\mu,r\right)|_{max}\nonumber\\
&=&V'\left(\omega,\mu,r_{0}\right)+\delta V''\left(\omega,\mu,r_{0}\right) + \frac{1}{2}\delta^{2}V^{(3)}\left(\omega,\mu,r_{0}\right)+\frac{1}{6}\delta^{3}V^{(4)}\left(\omega,\mu,r_{0}\right)\nonumber\\
&& +\frac{1}{24}\delta^{4}V^{(5)}\left(\omega,\mu,r_{0}\right) + \frac{1}{120}\delta^{5}V^{(6)}\left(\omega,\mu,r_{0}\right)+\frac{1}{720}\delta^{6}V^{(7)}\left(\omega,\mu,r_{0}\right),
\end{eqnarray}
where the primes and the superscript with the number in the parentheses denote $n-th$ derivatives with respect to $r$. By substituting Eqs.~(\ref{vmu}) and (\ref{rmu}) in Eq.~(\ref{rmaxcondi}), one can see that $r_{0}$ can be obtained by solving the zeroth order equation. Then the other values of $r_{i}$ can be obtained by solving the equations order by order as the following
\begin{eqnarray}
0&=&\mu\left[V_{1}'\left(\omega,r_{0}\right) + r_{1}V_{0}''\left(\omega,r_{0}\right)\right],\nonumber\\
0&=&\mu^{2}\left[V_{2}'\left(\omega,r_{0}\right)+r_{2}V_{0}''\left(\omega,r_{0}\right)+r_{1}V_{1}''\left(\omega,r_{0}\right) +\frac{1}{2}r_{1}^{2}V_{0}^{(3)}\left(\omega,r_{0}\right)\right],\nonumber\\
0&=&\mu^{3}\left[V_{3}'\left(\omega,r_{0}\right)+r_{3}V_{0}''\left(\omega,r_{0}\right)+r_{2}V_{1}''\left(\omega,r_{0}\right) +r_{1}V_{2}''\left(\omega,r_{0}\right)\right]\nonumber\\
&&+\mu^{3}\left[r_{1}r_{2}V_{0}^{(3)}\left(\omega,r_{0}\right)+\frac{1}{2}r_{1}^{2}V_{1}^{(3)}\left(\omega,r_{0}\right) +\frac{1}{6}r_{1}^{3}V_{0}^{(4)}\left(\omega,r_{0}\right)\right],\nonumber\\
&\vdots&.
\end{eqnarray}
As an example, by choosing the set of parameters as $\beta_{m}=0.5$, $c_{2}=-0.02/3$, and $\lambda=l+1=6$, the coefficients $r_{i}$ can be expressed in terms of $\omega$ up to the sixth order as follows
\begin{eqnarray}
r_{0}&=&1.3571,\nonumber\\
r_{1}&=&-\frac{0.2894}{\omega},\nonumber\\
r_{2}&=&0.5918 +\frac{0.1913}{\omega^2},\nonumber\\
r_{3}&=&\frac{0.0809}{\omega}-\frac{0.1290}{\omega^{3}},\nonumber\\
r_{4}&=&0.5464 -\frac{0.8466}{\omega^2} +\frac{0.0853}{\omega^4},\nonumber\\
r_{5}&=&\frac{0.6796}{\omega}+\frac{1.5032}{\omega^3}-\frac{0.0537}{\omega^5},\nonumber\\
r_{6}&=&0.3560 -\frac{0.9991}{\omega^2}-\frac{1.9547}{\omega^4}+\frac{0.0312}{\omega^6}.
\end{eqnarray}
By keeping $r_{max}$ up to the sixth order of $\mu$, we can evaluate the quantities in Eq.~(\ref{sscoes}), then substitute them into Eqs.~(\ref{absor}) and (\ref{ss}), allowing the greybody factor to be obtained. One can also check that when $\mu=0$, only the zeroth order survives, which infers the massless fermions case. It is important to note that we need to keep the series expansion up to the sixth order of $\mu$ in order to maintain the efficiency in the further expansion of the WKB approximation.

\subsubsection{Low energy approximation}
For low energy limit WKB (LWKB), where $\omega^{2}\ll V_{max}(\omega, m, r)$, the distance between the turning points denoted by $r_1$ and $r_2$ are large. Therefore, the wave solution in each region for the WKB approximation will significantly differ from one for the intermediate energy approximation. By keeping the condition $\omega^{2}\ll V_{max}(\omega, m, r)$ when performing the WKB expansion, the greybody factor is given by \cite{Cho:2004wj}
\begin{equation}
T=\exp\left({-2\int_{r_{1*}}^{r_{2*}}  dr_{*} \sqrt{V(\omega,m,r)-\omega^{2}}}\right).
\end{equation}
By considering small mass limit as done in the intermediate approximation, one has $V(\omega,m,r)\approx V(r)$. By using the coordinate transformation in Eq.~(\ref{ttcoord}), the above equations can be written as
\begin{equation}
T\approx\exp\left({-2\int_{r_{1}}^{r_{2}}  dr  \frac{1}{f(r)} \sqrt{V(r)-\omega^{2}}}\right).
\end{equation}
Note that the turning points $r_{1}$ and $r_{2}$ satisfy the relation $V(r_{1})=V(r_{2})=\omega^{2}$. By imposing the LWKB condition, $\omega^{2}\ll V_{max}(r)$, we can simplify the above equation as
\begin{equation}\label{transle}
T\approx\exp{\left[-2\int_{r_{1}}^{r_{2}}  dr  \left(\frac{\sqrt{V(r)}}{f(r)} -\frac{\omega^{2}}{2f(r) \sqrt{V(r)}}\right)\right]}.
\end{equation}
For the given $\omega$ which satisfies $\omega^{2}\ll V_{max}(r)$, we can solve the corresponding turning points $r_{1}$ and $r_{2}$. Then the greybody factor can be obtained numerically by performing integration in Eq.~(\ref{transle}). Note that the obtained result is just for one value of $\omega$, so that in this case, we have to perform the integration ``point by point" numerically.
\subsection{Results}
\subsubsection{massless fermion}
For the methods presented in the previous subsection, the zeroth order expansion of $\mu$ with $r_{max}=r_{0}$ is the massless case. For the massless case, we have three parameters to specify; $\beta_m$, $c_2$ and $l$. Therefore, in numerical results, we fix two of them and vary the other one to see how the parameter affects the greybody factor.

\begin{figure}
\begin{subfigure}{0.4\textwidth}
\includegraphics[width=\textwidth]{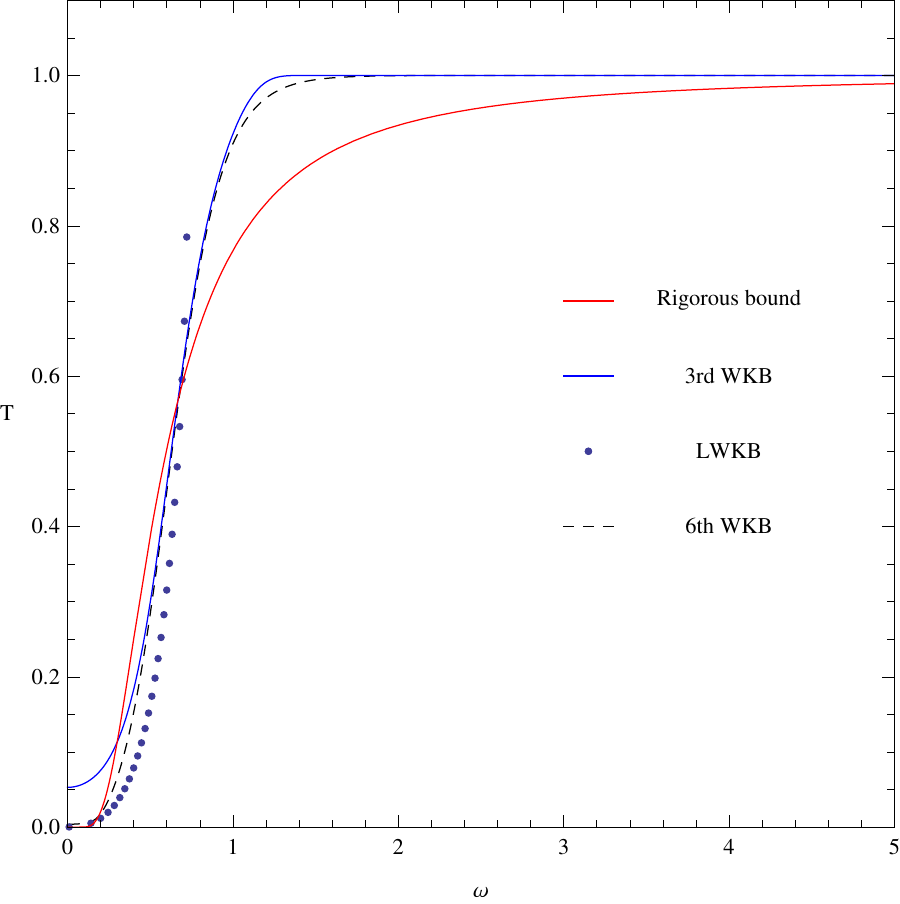}
\caption{$l=0$.}\label{b05c2l0m0}
\end{subfigure}
\begin{subfigure}{0.4\textwidth}
\includegraphics[width=\textwidth]{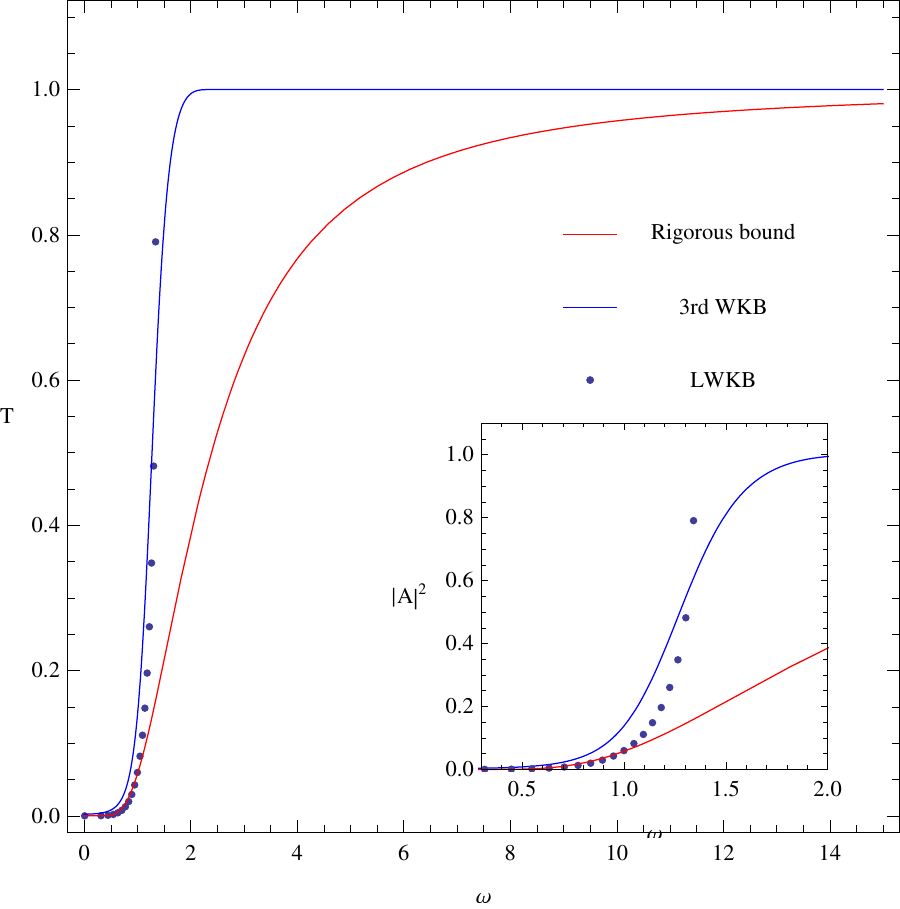}
\caption{$l=1$.}\label{b05c2l1m0}
\end{subfigure}\\
\begin{subfigure}{0.4\textwidth}
\includegraphics[width=\textwidth]{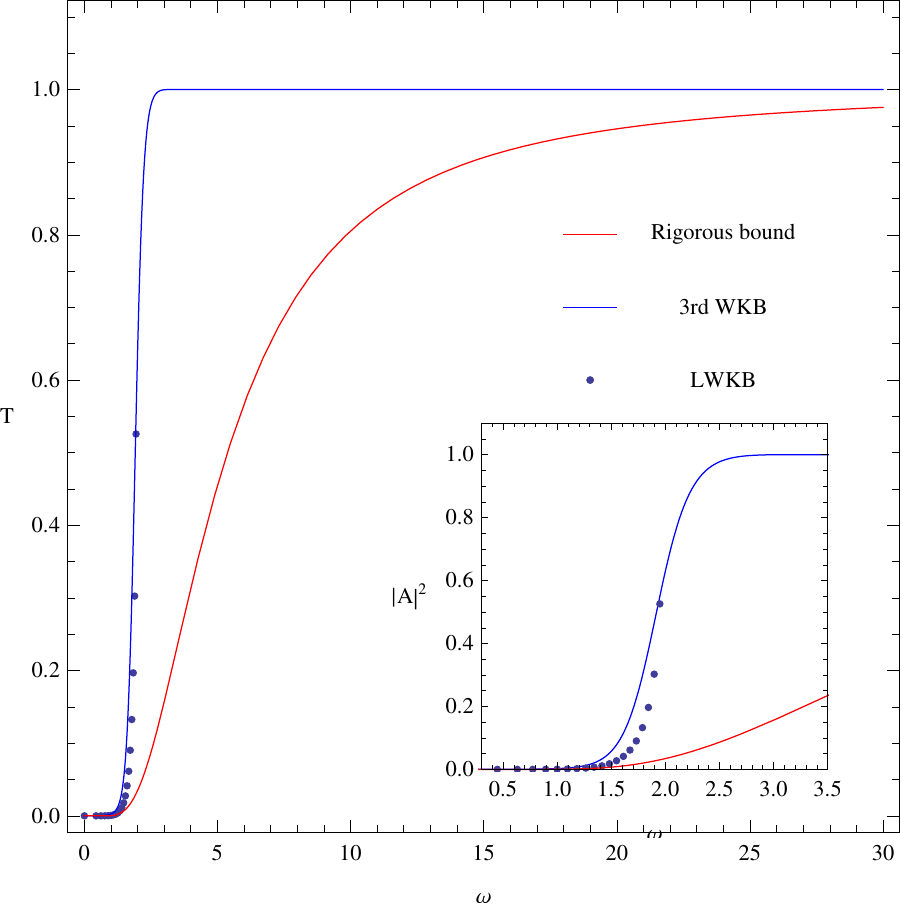}
\caption{$l=2$.}\label{b05c2l2m0}
\end{subfigure}
\begin{subfigure}{0.4\textwidth}
\includegraphics[width=\textwidth]{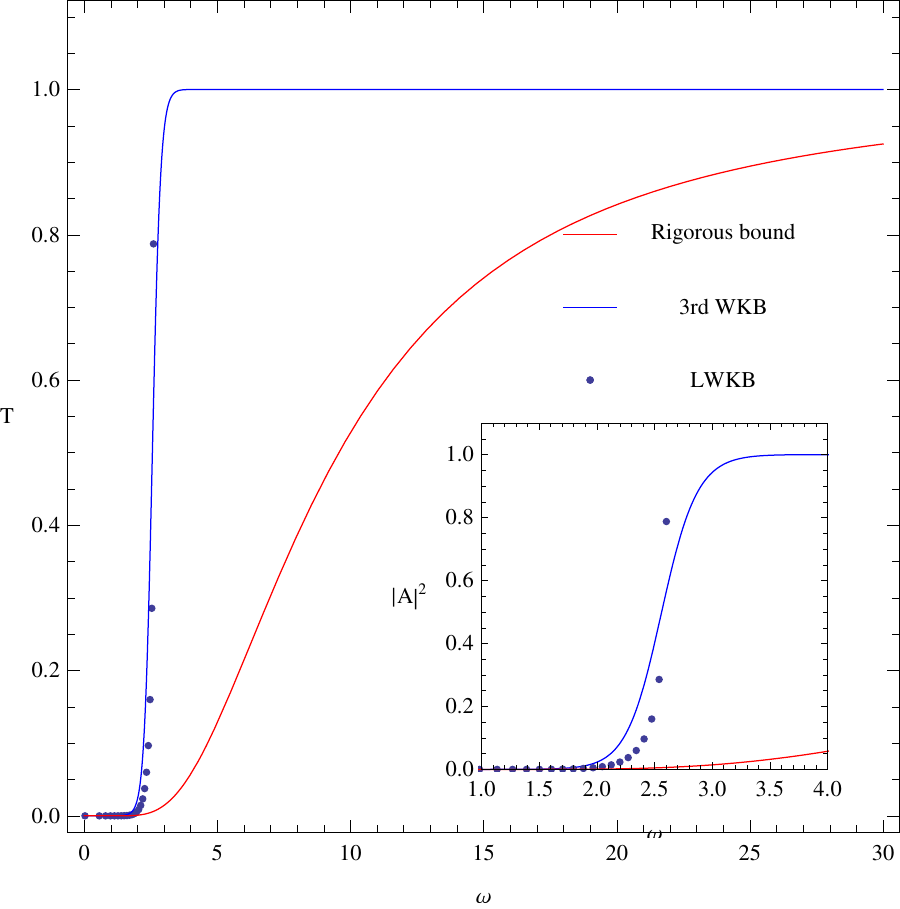}
\caption{$l=3$.}\label{b05c2l3m0}
\end{subfigure}
\caption{The greybody factor for massless Dirac particle with $\beta_{m}=0.5$, $c_{2}=-2/300$ and change $l$.}\label{b05c2m0}
\end{figure}

First let us consider the variation of $l$ by setting parameters $\beta_{m}=0.5$, $c_{2}=-2/300$, where $l$ is chosen as $l=0,\ 1,\ 2,\ 3$. This is equivalent to the spin -1/2 angular eigenvalue $\lambda=1,\ 2,\ 3,\ 4$. The numerical results for the greybody factor are illustrated in Fig.~\ref{b05c2m0}. The solid red lines represent the rigorous bound, the solid blue lines represent the 3rd order WKB results, and the blue dots represent the LWKB results. As illustrated in the left panel of Fig.~\ref{fig:VT-l}, the effective potential is higher as $l$ increases. The Dirac particles are necessary to include higher energy to transmit the effective potential. As a result, the greybody factor profile shifts to a larger $\omega$ region as $l$ increases. For the $l=0$ case, there is an inconsistency in the rigorous bound and the WKB methods, since the rigorous bound is the analytical lower bound, and then WKB results should be higher than the bound as presented in the $l=1,\ 2,\ 3$ cases (where it is more clear to check in the zone-in sub-figure). For this inconsistency, we also check the accuracy of the WKB method by performing the calculation up to the 6th order of the WKB corrections as found in Fig.~\ref{b05c2l0m0}. As a result, the inconsistency is still be present. As discussed in \cite{Konoplya:2003ii}, for the $l=0$ case, the eikonal formula in Eq. (\ref{ss}) does not give a good estimation for the greybody factor, except for large $\omega$ and the result does not improve the accuracy significantly with the increase in the WKB order. This is one of the disadvantages of the WKB method. From Fig.~\ref{b05c2m0}, one can see that the rigorous bound at a high value of $l$ is much lower than the results from the WKB method. It is still valid, but it may not be useful since some tiny effect may be lost. Therefore, in this state, one can argue that it is useful to use the rigorous bound method for the low potentials and use the WKB method for the high potentials.

\begin{figure}
\begin{subfigure}{0.4\textwidth}
\includegraphics[width=\textwidth]{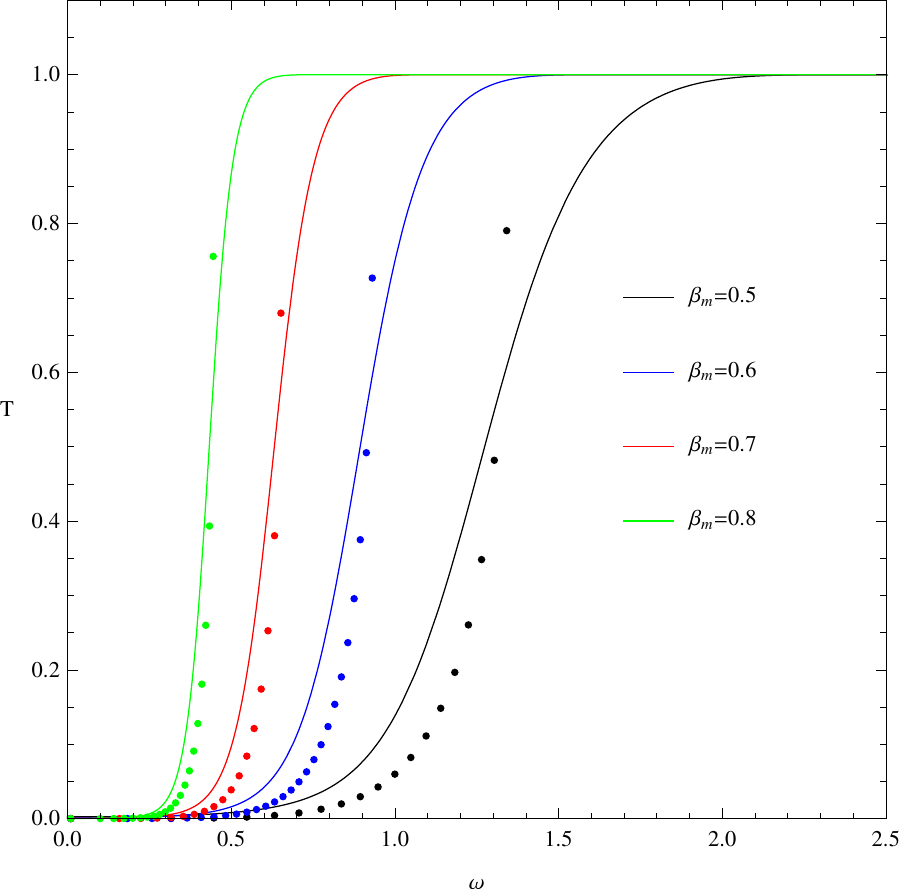}
\caption{The comparison for Greybody factor with $c_{2}=-2/300$, $l=1$, and $\beta_{m}=0.5,\ 0.6,\ 0.7,\ 0.8$.}\label{b5678llc2}
\end{subfigure}
\begin{subfigure}{0.4\textwidth}
\includegraphics[width=\textwidth]{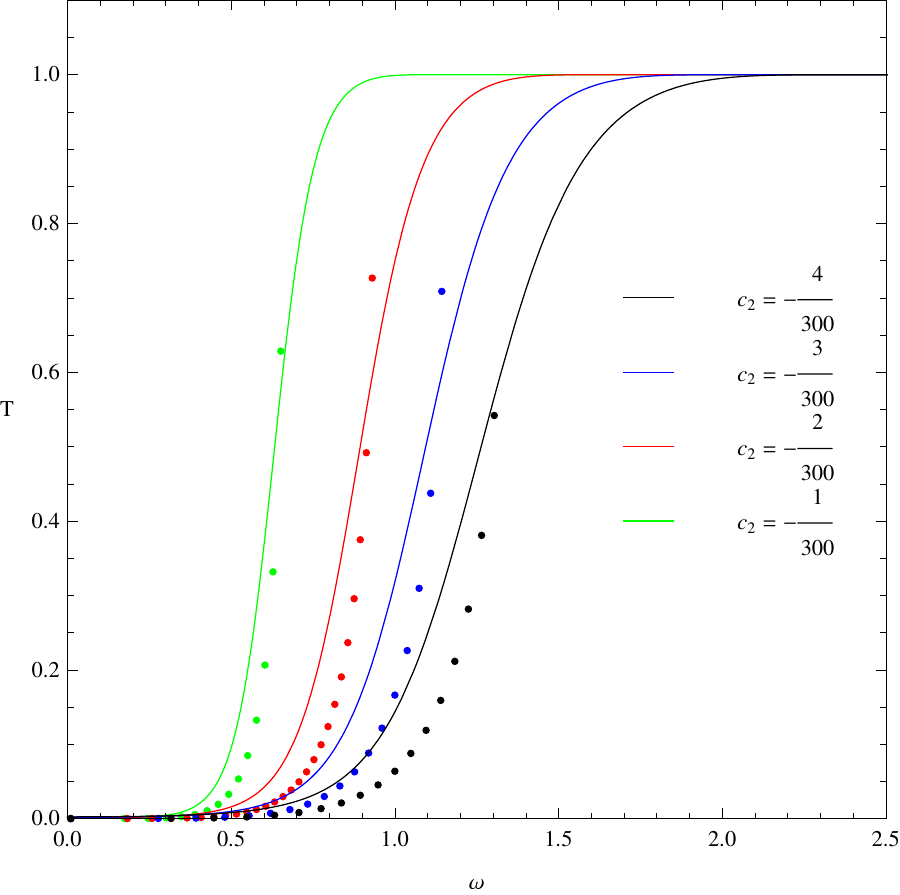}
\caption{The comparison for greybody factor with $\beta_{m}=0.6$, $l=1$, and $c_{2}=-1/300,\ -2/300,\ -3/300,\ -4/300$.}\label{b6c1234l1}
\end{subfigure}
\caption{The greybody factor for massless Dirac particle corresponding to the effective potential in Fig.~\ref{fig:VT-bm} and Fig.~\ref{fig:VT-c2}.}\label{comparebc2}
\end{figure}

Next, we continue our study for the cases of fixing $c_{2}=-2/300$, $l=1$, and varying $\beta_{m}=0.5,\ 0.6,\ 0.7,\ 0.8$, which corresponds to the effective potential in Fig.~\ref{fig:VT-bm}. The results are presented in Fig.~\ref{b5678llc2}, where the solid line represents the 3rd order WKB results and the dots represent the LWKB results. The greybody factor curve is shifted to a larger $\omega$ region when $\beta_{m}$ increases. This satisfies the behavior of the effective potential, which is higher when $\beta_{m}$ increases. Note that the corresponding $\omega$ is approximated as $\omega^{2}\approx V_{max}$ for the greybody factor $T\approx 0.5$, which will be considered as the reliable results for the study of greybody factor using the WKB approximation. For the case of fixing $\beta_{m}=0.6$, $l=1$ and varying $c_{2}=-1/300,\ -2/300,\ -3/300,\ -4/300$, the results are presented in Fig.~\ref{b6c1234l1} and the corresponding effective potential is presented in Fig.~\ref{fig:VT-c2}. Note that the WKB results are consistent with the rigorous bound for the cases listed above, which covers most of the cases except some of the $l=0$ ones. Therefore, we have omitted the rigorous bound result in Fig.~\ref{comparebc2} as they are already presented in Figs.~\ref{fig:VT-bm} and \ref{fig:VT-c2}.


\subsubsection{massive fermion}
\begin{figure}
\begin{subfigure}{0.4\textwidth}
\includegraphics[width=\textwidth]{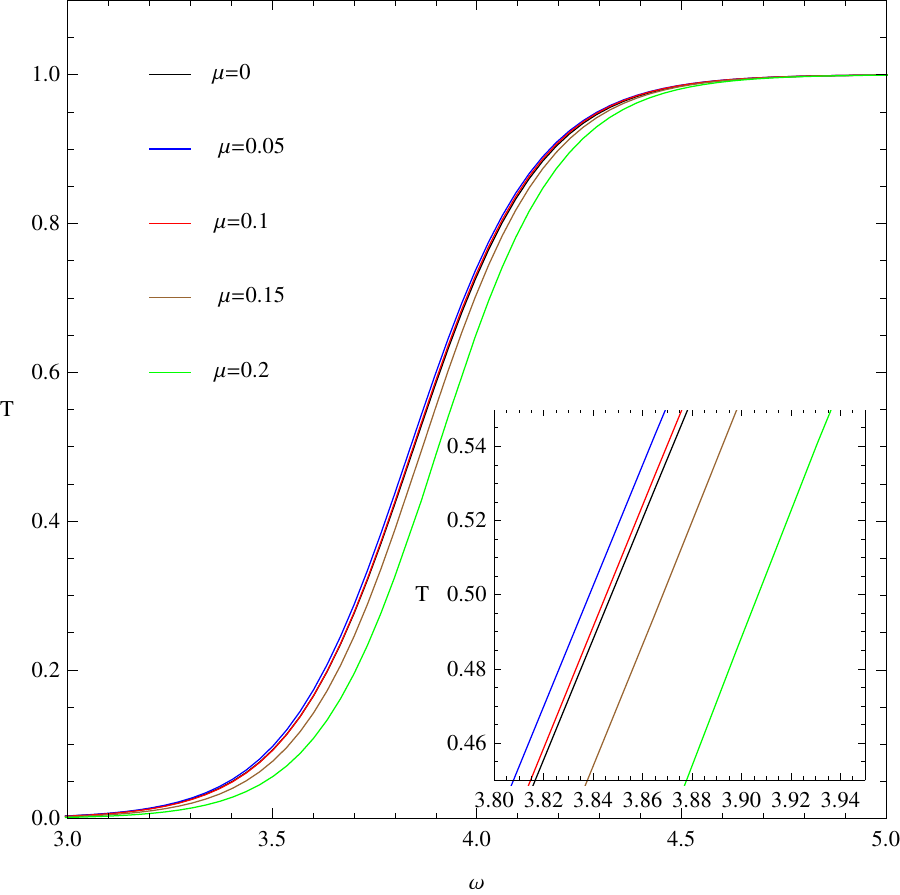}
\caption{$\mu=0,\ 0.05,\ 0.1,\ 0.15,\ 0.2$.}\label{b05c2l5mc}
\end{subfigure}
\begin{subfigure}{0.4\textwidth}
\includegraphics[width=\textwidth]{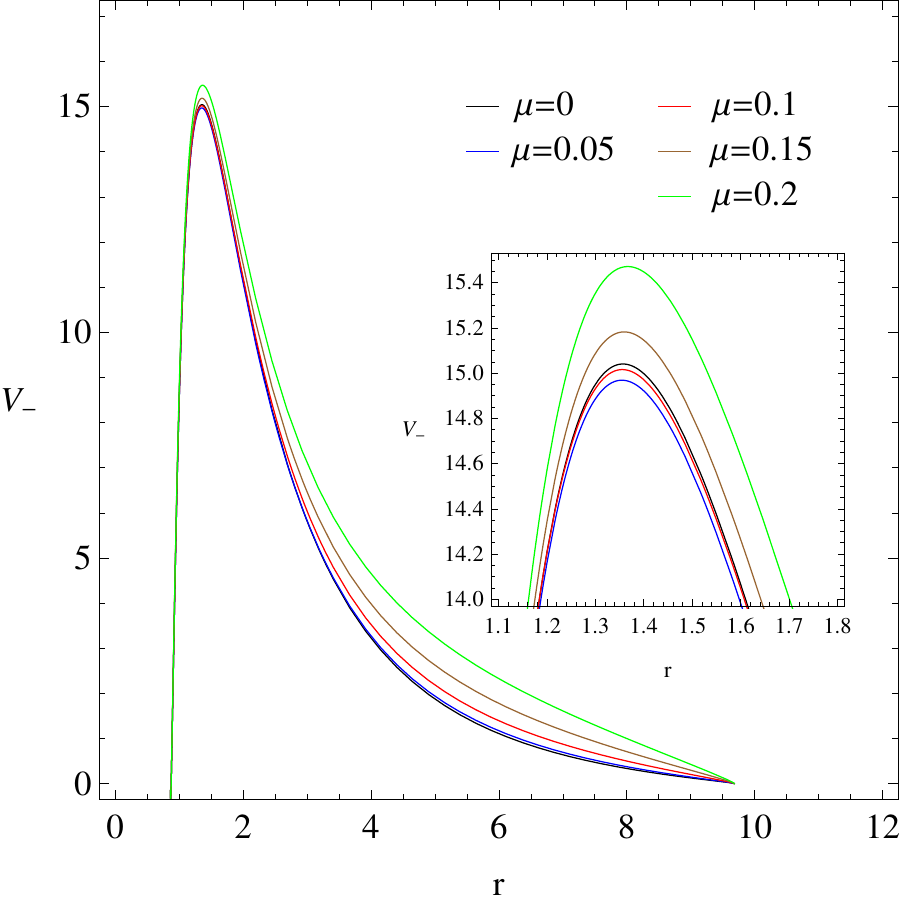}
\caption{Effective potential for $\omega=3.8$.}\label{b05c2l5mcv}
\end{subfigure}
\caption{The greybody factor and the effective potential for massive Dirac particle with $\beta_{m}=0.5$, $c_{2}=-2/300$, and $l=5$.}\label{b05c2l5m}
\end{figure}

\begin{figure}
\begin{subfigure}{0.4\textwidth}
\includegraphics[width=\textwidth]{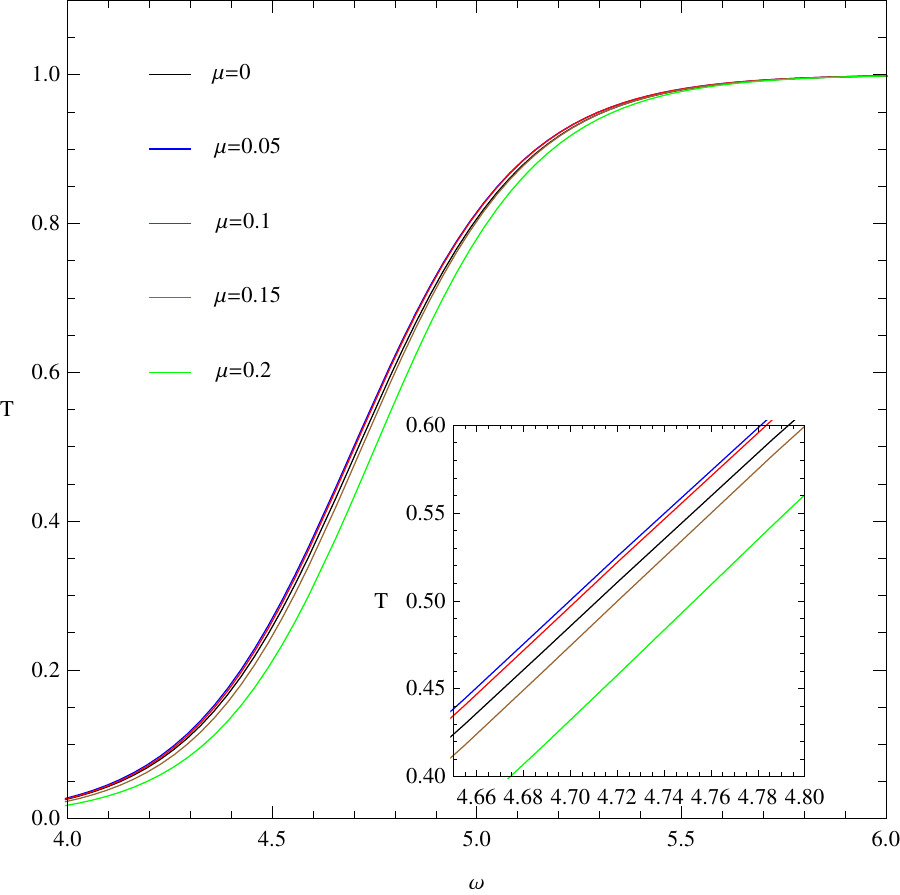}
\caption{$\mu=0,\ 0.05,\ 0.1,\ 0.15,\ 0.2$.}\label{b05c3l5mc}
\end{subfigure}
\begin{subfigure}{0.4\textwidth}
\includegraphics[width=\textwidth]{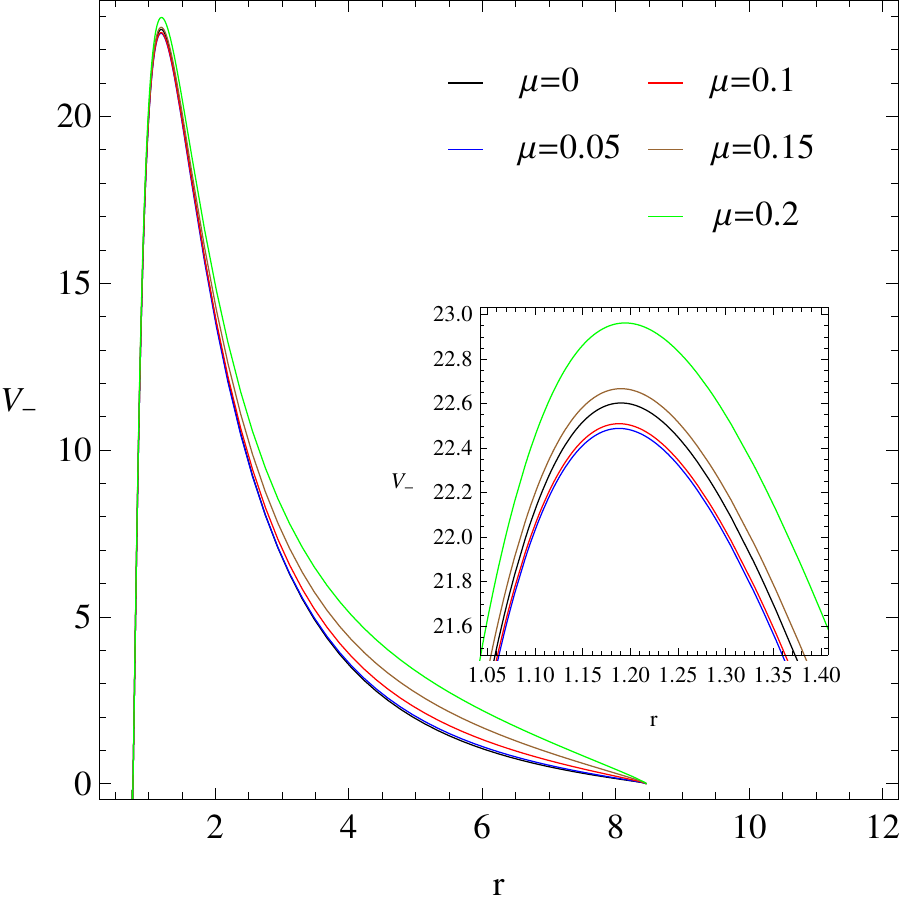}
\caption{Effective potential for $\omega=4.7$.}\label{b05c3l5mcv}
\end{subfigure}
\caption{The greybody factor and the effective potential for massive Dirac particle with $\beta_{m}=0.5$, $c_{2}=-3/300$, and $l=5$.}\label{b05c3l5m}
\end{figure}

\begin{figure}
\begin{subfigure}{0.4\textwidth}
\includegraphics[width=\textwidth]{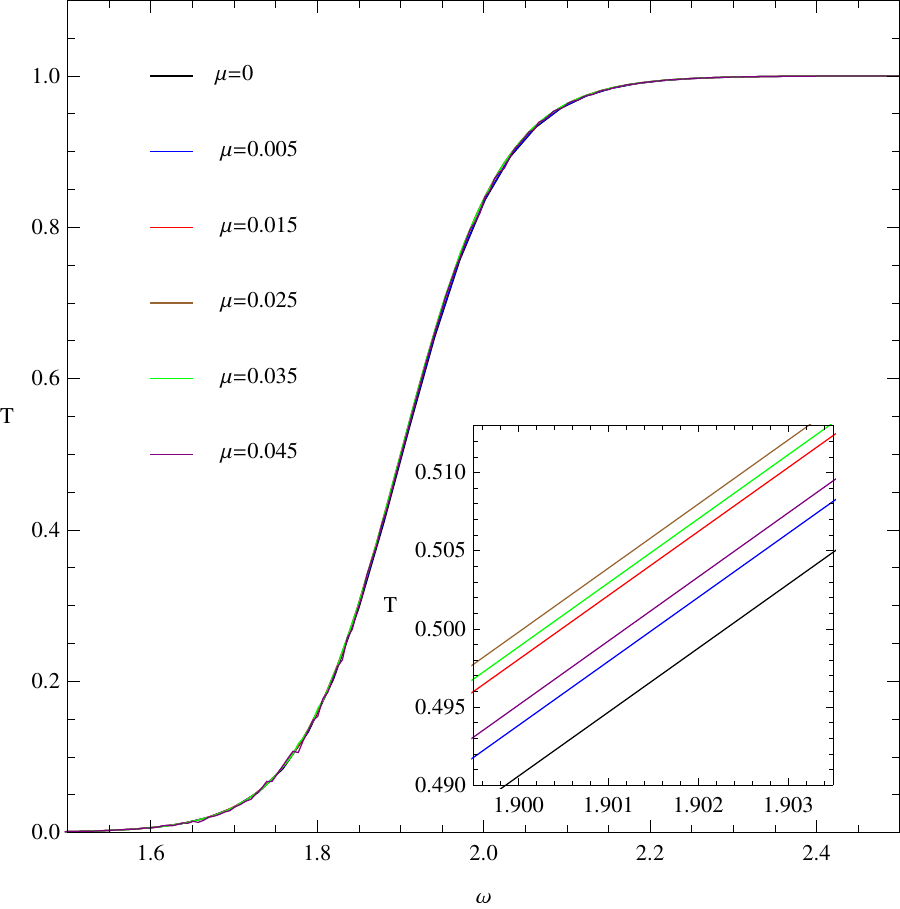}
\caption{$\mu=0,\ 0.005,\ 0.015,...,\ 0.045$.}\label{b07c2l5mc}
\end{subfigure}
\begin{subfigure}{0.4\textwidth}
\includegraphics[width=\textwidth]{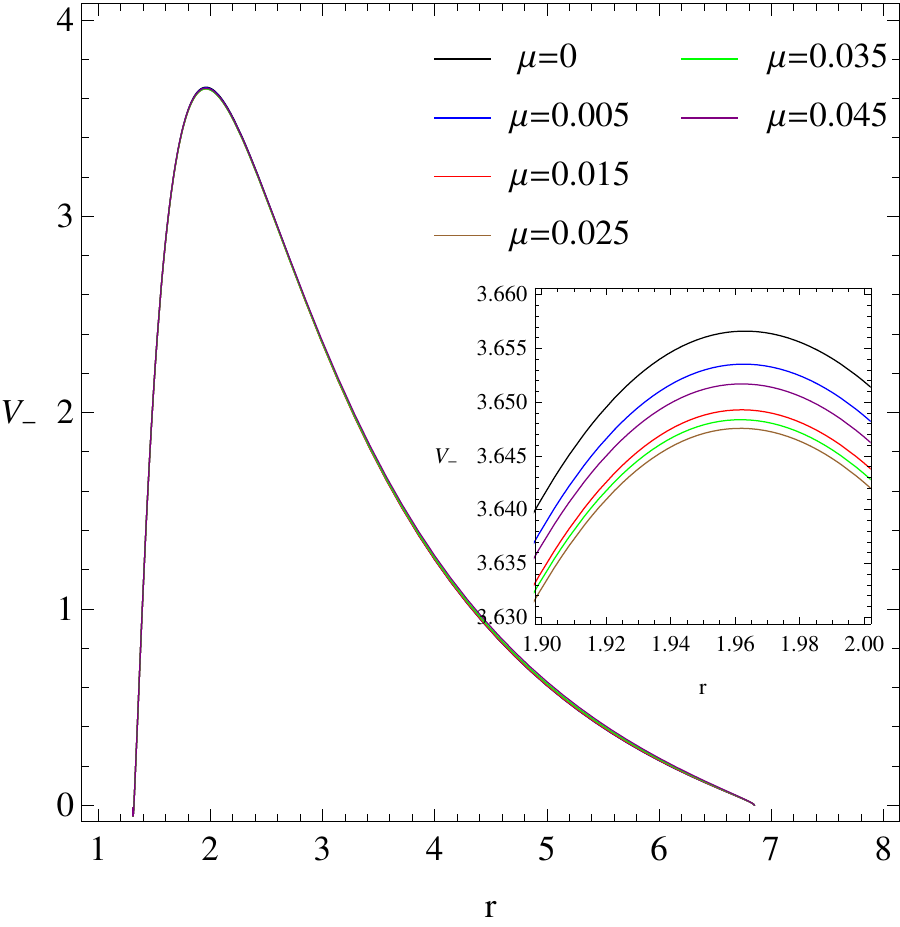}
\caption{Effective potential for $\omega=1.901$.}\label{b07c2l5mcv}
\end{subfigure}
\caption{The greybody factor and the effective potential for massive Dirac particle with $\beta_{m}=0.7$, $c_{2}=-2/300$, and $l=5$.}\label{b07c2l5m}
\end{figure}

For the massive case, we need to consider the full expressions discussed in Sec.~(\ref{IEA}) by evaluating until the 6th order of $\mu$. In order to clearly make our presentation, we must consider the larger $l$ cases. These cases correspond to the stronger effective potentials, and then it becomes easier to observe the difference when varying $\mu$ as presented in Fig.~\ref{b05c2l5mcv}, Fig.~\ref{b05c3l5mcv}, and Fig.~\ref{b07c2l5mcv}. Note that we have only considered the intermediate energy WKB results, while the low energy approximation results are analogous to the massless one.

In Fig.~\ref{b05c2l5mc}, we present the greybody factors for $\beta_{m}=0.5$, $c_{2}=-2/300$, $l=5$ and for the varying $\mu=0,\ 0.05,\ 0.1,\ 0.15,\ 0.2$. Note that the Dirac mass $m=(l+1)\mu$ in these cases is equivalent to $m=0,\ 0.3,\ 0.6,\ 0.9,\ 1.2$. One may observe that with varying $\mu$, the greybody factors do not explicitly change much compared to the $\mu=0$ case. The crucial behaviour in this case is that there is a critical point $\mu_c \sim 0.05$. From the zone-in sub-figure in Fig.~\ref{b05c2l5mc}, we can observe that when fixing $T=0.5$, the corresponding transmission energy $\omega$ decreases (shift to the left hand side) when $\mu$ increases from $0$ to $0.05$, then starts to increase (shift to the right hand side) when $\mu$ increases from $0.05$ to $0.2$. This behaviour can be found to satisfy the corresponding effective potentials as shown in sub-figure in Fig.~\ref{b05c2l5mcv}. The behavior for the maximum effective potential is consistent with the greybody factor when varying $\mu$. Note that for the effective potential shown in Fig.~\ref{b05c2l5mcv}, we use $\omega=4.7$, which is an approximate value of $\omega$ for $T\sim0.5$. This choice satisfies the intermediate condition, $\omega^2 \sim V_{max}$.

\begin{figure}
\begin{subfigure}[t]{0.4\textwidth}
\includegraphics[width=\textwidth]{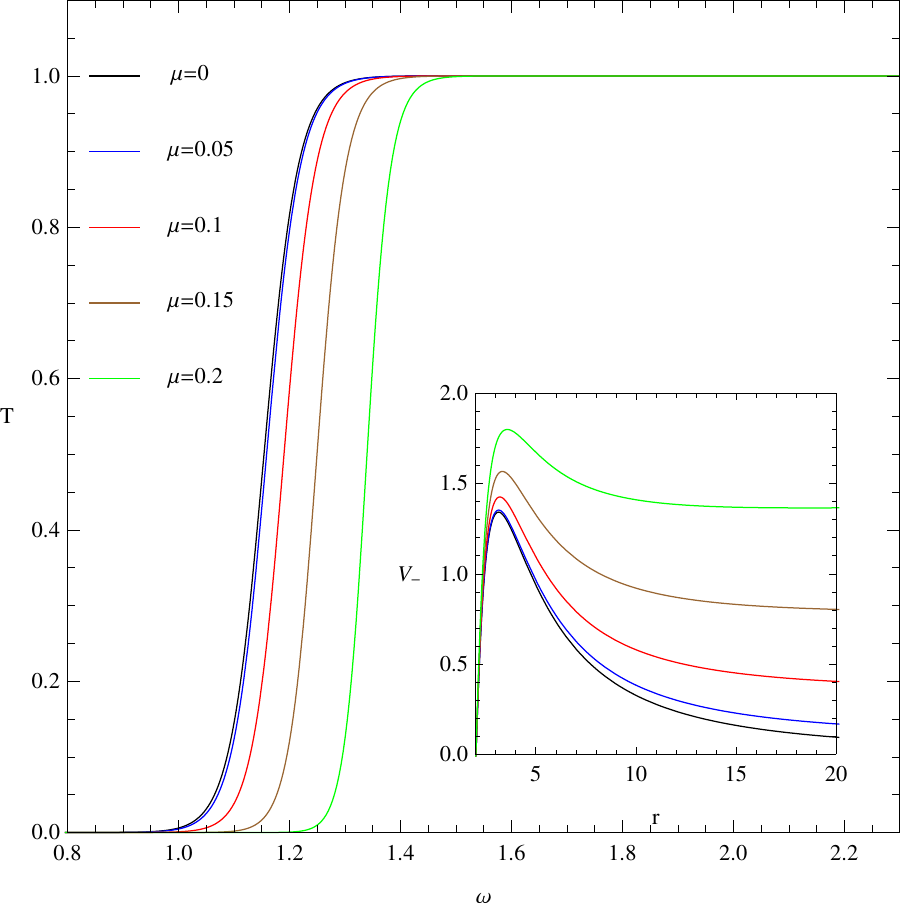}
\caption{Schwarzschild cases with $M=1$, $l=5$, and $\mu=0,...0.2$.}\label{Schwl5m}
\end{subfigure}
\begin{subfigure}[t]{0.403\textwidth}
\includegraphics[width=\textwidth]{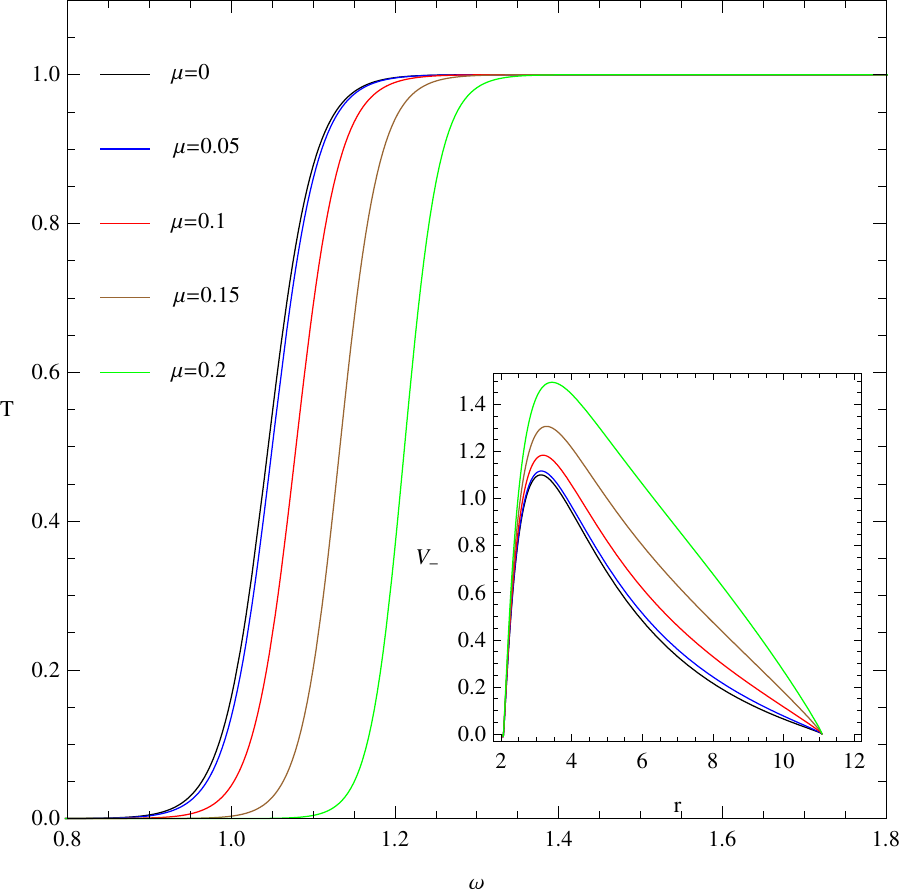}
\caption{Schwarzschild-dS cases with $M=1$, $\Lambda=0.02$, $l=5$, and $\mu=0,...0.2$.}\label{SdSl5m}
\end{subfigure}
\caption{The greybody factors and the effective potentials for massive Dirac particle in Schwarzschild and Schwarzschild-dS spacetimes.}\label{SSdS}
\end{figure}

In Fig.~\ref{b05c3l5m}, we present the greybody factors for $\beta_{m}=0.5$, $c_{2}=-3/300$, $l=5$ and for the varying $\mu=0,\ 0.05,\ 0.1,\ 0.15,\ 0.2$, which is equivalent to $m=0,\ 0.3,\ 0.6,\ 0.9,\ 1.2$. A similar behaviour as for the previous case is found. There exists the critical point $\mu_c \sim 0.05$ such that the greybody factors decrease when $\mu$ increases from $0$ to $0.05$, then increase when $\mu$ increases from $0.05$ to $0.2$, respectively. In Fig~\ref{b07c2l5m}, we present a case with a smaller set of effective potentials, while fixing parameters as $\beta_{m}=0.7$, $c_{2}=-2/300$, $l=5$. Again, it is found that there exists the critical point around $\mu_c \sim 0.025$. This still satisfies the behaviour of the effective potential in the sub-figures of Fig.~\ref{b07c2l5mcv}. To summarize the findings of this stage, for the massive Dirac field in dRGT black hole when fixing parameters $\beta_{m}$, $c_{2}$ and $l$, the behaviour of the greybody factors decrease when we increase the Dirac mass parameter $\mu$ (or $m$) from massless one, then a critical point with specific $\mu_{c}$ exists and the behaviour of the greybody factors increase after the critical point. The critical point of $\mu_{c}$ may not be able to be evaluated explicitly from the numerical processes, but corresponds to the same critical point of the maximum effective potentials. It is worth to note that the critical point of $\mu_{c}$ does not appear in similar studies of Schwarzschild and Schwarzschild-dS spacetimes. The results are illustrated in Fig.~\ref{SSdS}. From this figure, we set the black hole mass $M=1$, the angular parameter $l=5$, and the Dirac mass parameter $\mu=0,\ 0.05,\ 0.1,\ 0.15,\ 0.2$ for the Schwarzschild black hole in Fig.~\ref{Schwl5m}. For the Schwarzschild-dS black hole, we use the same settings for $M,\ l,\ \mu$ and set the cosmological constant $\Lambda=0.02$ as shown in Fig.~\ref{SdSl5m}. The corresponding effective potential is also presented in the sub-figures, respectively.

It is worthwhile to note that there is a limit of mass parameter $\mu$ in computational calculation. For this limit, the computational results give fluctuation as shown in Fig. \ref{b07c2l5msuc}, in Appendix \ref{AppB}. We observe that the limit depends on the shape of the potential. Actually, it seems like if the potential is lower, the limit of mass parameters is lower. For example, the limit becomes lower for increasing parameter $\beta_m$. By setting $l=5, c_2=-1/300 $, we obtain the limit as $\mu \lesssim 0.4$ for $\beta_m = 0.5$, corresponding to higher potential, while the limit becomes $\mu \lesssim 0.05 $ for parameter setting $ \beta_m = 0.7$, corresponding to lower potential. We investigate this issue by varying three parameters $l, c_2, \beta_m$ as shown in Table \ref{Tab1}, in Appendix \ref{AppB}. These results agree with the investigations in literature, which suggests that the WKB approximation cannot work well for lower multipole $l$, corresponding to lower potential. Since our method uses the expansion by requiring the mass parameter to be small, the corrections from the higher order of $\mu$ will influence the WKB approximation in case the parameter $\mu$ is not small enough.

\section{Conclusion}\label{con}
In this paper, we investigated the greybody factor from the Dirac fields on black holes in dRGT massive gravity theory. The greybody factor is obtained using two methods, the rigorous bound and the WKB. For the rigorous bound method, it provides us with a useful way to qualitatively analyze the behaviour of the greybody factor. We have separated our investigation into two parts; the massless and the massive cases. For the massless case, the crucial contribution to the greybody factor directly depends on the distance between two horizons as shown in Eq. \eqref{Tb-massless}. As a result, the greybody factor which is equivalent to the transmission coefficient significantly depends on the shape of the potential. If the potential is small, there is more probability for the Dirac particle to transmit through the black hole, and then the greybody factor is higher.

For the massive case, the full expression for the greybody factor bound is complicated and also difficult to analyze qualitatively. Therefore, we use two approximated expressions to analyze how the greybody factor depends on the mass of the Dirac field. For the first expression in Eq. \eqref{Tapp1}, we found that the Dirac field with heavier mass tends to be more difficult to transmit through the black hole, therefore, the greybody factor is lower. By comparing the results to one from the full expression, we found that even though the bound is still valid, it does not provide a significant behaviour of the greybody factor at very low mass. Specifically, the greybody factor will increase as the mass increase at some range for very small masses as shown in Fig. \ref{fig:Tmufullapp}. We then use the second approximated expression to find the critical point for which the greybody factor is maximized as seen in Eq. \eqref{mut}. We also found that the effect of the shape of the potential on the greybody factor is still the same as for the massless case. However, the critical mass $\mu_c$ dose not provide the lowest peak of the potential. This may be one of the disadvantages of the rigorous bound method. The bound is still valid, but some tiny effect may be lost. Moreover, for large $\lambda$, the bound is much lower than the exact value obtained in other methods.

It is interesting that our qualitative analysis of the massive case is quite general. Therefore, it is useful to apply the strategy performed in the Dirac field case to other cases such as massive scalar and massive vector fields. We leave this investigation to further works.

For the WKB method, the regular procedure may not be applied since the potential depends on both mass and energy of the Dirac field. Therefore, we divide our investigation into two parts; the intermediated energy $\omega^2 \sim V_{max}$ and low energy $\omega^2 \ll V_{max}$. For the intermediated energy case, we apply the series expansion of the potential in every step of computation of the WKB series by keeping the mass parameter $\mu$ small. As a result, the zero order of the series corresponds to the massless case. For the low energy case, the expression for the greybody factor can be obtained. However,  we need to fix the energy $\omega$ in order to compute the greybody factor, then the evaluation can be performed point by point.

The resulting greybody factor from the WKB method shows that the WKB method does not work well for low multipole $\lambda$, which corresponds to low potential. This also agrees with other investigations in literature. In this case, it is worthwhile to use the rigorous bound method. For the high multipole case, the WKB method provides greybody factor with sufficient precision. One can see a tiny effect of the mass parameter on the greybody factor, which is inferred from the behaviour of the potential. This is not trivial for the rigorous bound method. In this case, it is better to use the WKB method compared to the rigorous bound method.

From observational points of view, we may still be far away from detecting the relevant spectrum of the greybody factor. However, the existence of the critical mass may shed light on the connection between the theoretical prediction and the observation, since the maximum value of the greybody factor at the critical mass may provide clues of possible ways to detect the spectrum of the greybody factor.

It is important to note that our results are valid for small masses of the Dirac field. Actually, it is assumed that if there are no backreactions, then the black hole is stable. For large masses, the black hole may not be stable and the supperradiance may occur. We also leave this investigation to further works.

\section*{Acknowledgement}
This project was funded by the Ratchadapisek Sompoch Endowment Fund, Chulalongkorn University (Sci-Super 2014-032), by a grant for the professional development of new academic staff from the Ratchadapisek Somphot Fund at Chulalongkorn University, by the Thailand Research Fund (TRF), and by the Office of the Higher Education Commission (OHEC), Faculty of Science, Chulalongkorn University (RSA5980038). PB was additionally supported by a scholarship from the Royal Government of Thailand. TN was also additionally supported by a scholarship from the Development and Promotion of Science and Technology Talents Project (DPST). PW was supported by the Thailand Research Fund (TRF) through grant no. MRG6180003. PB and PW were supported by SERB-DST, India for the ASEAN project IMRC/AISTDF/CRD/2018/000042.

\appendix
\section*{Appendix}
\section{Analytic solution for massive fermion}\label{AppA}
For the massive case, consider two inner horizons, where $f(r) > 0$. Then
\begin{eqnarray}
\int_{-\infty}^{\infty}\left|W^{2}\right|dr_{*} &=& \int_{r_{H}}^{R_{H}}\frac{1}{r^{2}}\frac{\left(\lambda^{2} + m^{2}r^{2}\right)^{2}}{\lambda^{2} + m^{2}r^{2} + \left(\lambda m/2\omega\right)f(r)}dr\nonumber\\
     &=& \int_{r_{H}}^{R_{H}}2\omega\frac{m^{4}r^{4} + 2\lambda^{2}m^{2}r^{2} + \lambda^{4}}{2\omega m^{2}r^{4} + 2\omega\lambda^{2}r^{2} + \lambda mr^{2}f(r)}dr.
\end{eqnarray}
In the dRGT BH model, we have
\begin{equation}
f(r) = 1 - \frac{2\tilde{M}}{\tilde{r}} + \alpha_{g}\left(c_{2}\tilde{r}^{2} - c_{1}\tilde{r} + c_{0}\right).
\end{equation}
Therefore,
\begin{eqnarray}
\int_{-\infty}^{\infty}\left|W^{2}\right|dr_{*} &=& \int_{r_{H}}^{R_{H}}2\omega\frac{m^{4}r^{4} + 2\lambda^{2}m^{2}r^{2} + \lambda^{4}}{2\omega m^{2}r^{4} + 2\omega\lambda^{2}r^{2} + \lambda mr^{2}\left[1 - 2\tilde{M}/\tilde{r} + \alpha_{g}\left(c_{2}\tilde{r}^{2} - c_{1}\tilde{r} + c_{0}\right)\right]}dr\nonumber\\
     &=& \int_{r_{H}}^{R_{H}}2\omega\frac{m^{4}r^{4} + 2\lambda^{2}m^{2}r^{2} + \lambda^{4}}{\left(2\omega m^{2} + \lambda m \alpha_{g}c_{2}\right)r^{4} - \lambda m \alpha_{g}c_{1}r^{3} + \left(2\omega\lambda^{2} + \lambda m + \lambda m \alpha_{g}c_{0}\right)r^{2} - 2\tilde{M}\lambda mr}dr.\nonumber
\end{eqnarray}
Consider the integrand
\begin{equation}
\frac{m^{4}r^{4} + 2\lambda^{2}m^{2}r^{2} + \lambda^{4}}{\left(2\omega m^{2} + \lambda m \alpha_{g}c_{2}\right)r^{4} - \lambda m\alpha_{g}c_{1}r^{3} + \left(2\omega\lambda^{2} + \lambda m + \lambda m \alpha_{g}c_{0}\right)r^{2} - 2\tilde{M}\lambda mr}\nonumber
\end{equation}
\begin{eqnarray}
&=& A + \frac{B}{r}\nonumber\\
 && + \frac{Cr^{2} + Dr + E}{\left(2\omega m^{2} + \lambda m \alpha_{g}c_{2}\right)r^{3} - \lambda m\alpha_{g}c_{1}r^{2} + \left(2\omega\lambda^{2} + \lambda m + \lambda m \alpha_{g}c_{0}\right)r - 2\tilde{M}\lambda m}.\label{integrand}
\end{eqnarray}
We obtain
\[
A = \frac{m^{3}}{2\omega m + \lambda\alpha_{g}c_{2}}, ~~B = -\frac{\lambda^{3}}{2\tilde{M}m}, ~~C = \frac{\lambda m^{4}\alpha_{g}c_{1}}{2\omega m + \lambda\alpha_{g}c_{2}} + \frac{\lambda^{3}\left(2\omega m + \lambda\alpha_{g}c_{2}\right)}{2\tilde{M}}
\]
\[
D = \frac{\lambda m^{2}\left(2\omega\lambda m + 2\lambda^{2}\alpha_{g}c_{2} - m^{2} - m^{2}\alpha_{g}c_{0}\right)}{2\omega m + \lambda\alpha_{g}c_{2}} - \frac{\lambda^{4}\alpha_{g}c_{1}}{2\tilde{M}}
\]
and
\begin{equation}
E = \frac{2\tilde{M}\lambda m^{4}}{2\omega m + \lambda\alpha_{g}c_{2}} + \frac{\lambda^{4}\left(2\omega\lambda + m + m\alpha_{g}c_{0}\right)}{2\tilde{M}m}.
\end{equation}
Thus,
\begin{eqnarray}
\int_{-\infty}^{\infty}\left|W^{2}\right|dr_{*} &=& \frac{2\omega m^{3}}{2\omega m + \lambda\alpha_{g}c_{2}}\left(R_{H} - r_{H}\right) - \frac{\omega \lambda^{3}}{\tilde{M}m}\ln\left|\frac{R_{H}}{r_{H}}\right|\nonumber\\
     && + 2\omega\int_{r_{H}}^{R_{H}}\frac{Cr^{2} + Dr + E}{\left(2\omega m^{2} + \lambda m \alpha_{g}c_{2}\right)r^{3} - \lambda m\alpha_{g}c_{1}r^{2} + \left(2\omega\lambda^{2} + \lambda m + \lambda m \alpha_{g}c_{0}\right)r - 2\tilde{M}\lambda m}\nonumber\\
     &=& \frac{2\omega m^{3}}{2\omega m + \lambda\alpha_{g}c_{2}}\left(R_{H} - r_{H}\right) - \frac{\omega \lambda^{3}}{\tilde{M}m}\ln\left|\frac{R_{H}}{r_{H}}\right|\nonumber\\
     && + \frac{2\omega}{2\omega m^{2} + \lambda m \alpha_{g}c_{2}}\int_{r_{H}}^{R_{H}}\frac{Cr^{2} + Dr + E}{\left(r - R_{1}\right)\left(r - R_{2}\right)\left(r - R_{3}\right)},
\end{eqnarray}
where $R_{1}$, $R_{2}$ and $R_{3}$ are roots of equation
\begin{equation}
r^{3} - \frac{\lambda m\alpha_{g}c_{1}}{2\omega m^{2} + \lambda m \alpha_{g}c_{2}}r^{2} + \frac{2\omega\lambda^{2} + \lambda m + \lambda m \alpha_{g}c_{0}}{2\omega m^{2} + \lambda m \alpha_{g}c_{2}}r - \frac{2\tilde{M}\lambda m}{2\omega m^{2} + \lambda m \alpha_{g}c_{2}} = 0.
\end{equation}
By the method of partial fraction, we obtain
\begin{equation}
\frac{Cr^{2} + Dr + E}{\left(r - R_{1}\right)\left(r - R_{2}\right)\left(r - R_{3}\right)} = \frac{F}{r - R_{1}} + \frac{G}{r - R_{2}} + \frac{H}{r - R_{3}},
\end{equation}
which can be rewritten as
\begin{equation}
Cr^{2} + Dr + E = F\left(r - R_{2}\right)\left(r - R_{3}\right) + G\left(r - R_{1}\right)\left(r - R_{3}\right) + H\left(r - R_{1}\right)\left(r - R_{2}\right).
\end{equation}
Substituting $r = R_{1}$, $r = R_{2}$ and $r = R_{3}$, we obtain
\begin{equation}
F = \frac{CR_{1}^{2} + DR_{1} + E}{\left(R_{1} - R_{2}\right)\left(R_{1} - R_{3}\right)}, ~~G = \frac{CR_{2}^{2} + DR_{2} + E}{\left(R_{2} - R_{1}\right)\left(R_{2} - R_{3}\right)} ~~\textrm{and}~~H = \frac{CR_{3}^{2} + DR_{3} + E}{\left(R_{3} - R_{1}\right)\left(R_{3} - R_{2}\right)}.
\end{equation}
Therefore,
\begin{equation}
\int_{r_{H}}^{R_{H}}\frac{Cr^{2} + Dr + E}{\left(r - R_{1}\right)\left(r - R_{2}\right)\left(r - R_{3}\right)} = F\ln\left|\frac{R_{H} - R_{1}}{r_{H} - R_{1}}\right| + G\ln\left|\frac{R_{H} - R_{2}}{r_{H} - R_{2}}\right| + H\ln\left|\frac{R_{H} - R_{3}}{r_{H} - R_{3}}\right|.
\end{equation}
Finally, we obtain
\begin{eqnarray}
\int_{-\infty}^{\infty}\left|W^{2}\right|dr_{*} &=& \frac{2\omega m^{3}}{2\omega m + \lambda\alpha_{g}c_{2}}\left(R_{H} - r_{H}\right) - \frac{\omega \lambda^{3}}{\tilde{M}m}\ln\left|\frac{R_{H}}{r_{H}}\right|\nonumber\\
     && + \frac{2\omega}{2\omega m^{2} + \lambda m \alpha_{g}c_{2}}\left(F\ln\left|\frac{R_{H} - R_{1}}{r_{H} - R_{1}}\right| + G\ln\left|\frac{R_{H} - R_{2}}{r_{H} - R_{2}}\right|\right.\nonumber\\
     && \left. + H\ln\left|\frac{R_{H} - R_{3}}{r_{H} - R_{3}}\right|\right).
\end{eqnarray}
From equation (\ref{TT}), the rigorous bound on greybody factor is given by
\begin{eqnarray}
T &\geq& \textrm{sech}^{2}\left(\frac{1}{2\omega}\left[W(R_{H}) - W(r_{H})\right] + \frac{m^{3}}{2\omega m + \lambda\alpha_{g}c_{2}}\left(R_{H} - r_{H}\right) - \frac{\lambda^{3}}{2\tilde{M}m}\ln\left|\frac{R_{H}}{r_{H}}\right|\right.\nonumber\\
      && + \frac{1}{2\omega m^{2} + \lambda m \alpha_{g}c_{2}}\left[F\ln\left|\frac{R_{H} - R_{1}}{r_{H} - R_{1}}\right| + G\ln\left|\frac{R_{H} - R_{2}}{r_{H} - R_{2}}\right|\right.\nonumber\\
      && \left.\left. + H\ln\left|\frac{R_{H} - R_{3}}{r_{H} - R_{3}}\right|\right]\right).
\end{eqnarray}

\section{Computatinal efficiency}\label{AppB}
For the computational efficiency of  the greybody factor of the massive Dirac particles in the Schwarzschild black hole \cite{Cho:2004wj}, the WKB formula is sufficient only for $m<\omega$ because the asymptotic behaviour of the effective potential goes to $m^{2}$, which means the effective potential includes a phase transition from the barrier-like potential to the step function-like potential when $m\simeq\omega$. One can see this from the sub-figure of Fig.~\ref{Schwl5m}. According to this behaviour, this constraint is not valid for the Schwarzschild-dS and the dRGT black hole cases since the effective potential is always zero at the cosmic horizon as illustrated in Fig.~\ref{SdSl5m} for Schwarzschild-dS, as well as all the effective potential plots for dRGT black holes present in this paper. However, in evaluating the greybody factor through a further expansion of the WKB approximation, a ``numerical constraint" still exist even though the effective potentials are confirmed to be barrier-like. We examine the constraint on the Dirac mass parameter $\mu$ case by case by varying $l=2,\ 3,\ 4,\ 5$, $\beta_{m}=0.5,\ 0.6,\ 0.7,\ 0.7$, and $c_{2}=-1/300,\ -2/300,\ -3/300,\ -4/300$. The results are shown in Tab.~\ref{Tab1}. In this table, the blank column represents the ``successful evaluations" that occur when $\mu<10^{-10}$, which is a nearly massless result. It is found that a larger $l$ provides a stronger effective potential, and then a stronger effective potential leads to a successful evaluation with larger $\mu$. However, it is not exactly true when comparing the cases of the effective potentials in the same order, for example, for the cases of $l=2$, $\beta_{m}=0.5$, $c_{2}=-2/300$ and  $l=5$, $\beta_{m}=0.6$, $c_{2}=-1/300$. This may occur from the fact that even though the potential gets higher by increasing the magnitude of $c_2$, the potential is also thinner as shown in Fig.~\ref{fig:VT-c2}. This can also be seen from Tab.~\ref{Tab1} where the parameter $c_2$ changes. When we compare more cases listed in the table, we find that the condition on $\mu$ is not based on a single but various phenomena, including the strength of the effective potentials, and the structure of metric elements, as well as some numerical error.

\begin{figure}
\includegraphics[width=0.5\textwidth]{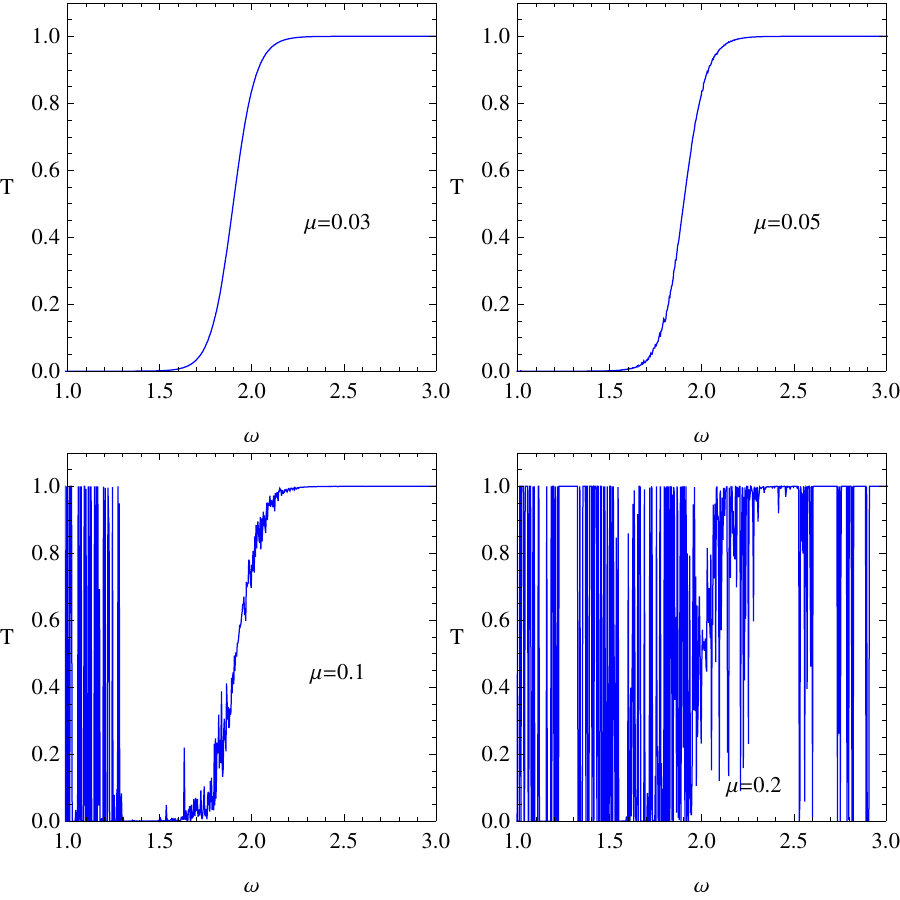}
\caption{Example of efficient areas for greybody factors with the parameters $\beta_{m}=0.7$, $c_{2}=-2/300$, $l=5$, and $\mu=0.03,\ 0.05,\ 0.1,\ 0.2 $.}\label{b07c2l5msuc}
\end{figure}

Lastly, we explain more about how we select for ``successful evaluations" and then provide the limit for the parameter $\mu$. In Fig.~\ref{b07c2l5msuc}, we show the case with $\beta_{m}=0.7$, $c_{2}=-2/300$, $l=5$ as an example and for varying $\mu$, where $\mu=0.03,\ 0.05,\ 0.1,\ 0.2$. One can observe that the numerical result starts displaying an irregular behaviour with $\mu=0.2$, such that the locus of greybody factor is not clear. For the case of $\mu=0.1$, the locus becomes clearer, but still includes some fluctuation. The shape then becomes more stable for the $\mu=0.05$ and $\mu=0.03$ cases. The upper two plots are what we call a ``successful evaluation", setting a constraint on $\mu\lesssim0.05$, as shown in Tab.~\ref{Tab1}.

\begin{table}
\centering
\caption{The efficient discussion.}
\begin{tabular}{| c | c | c | c | c || c | c | c | c | }
\hline
& \multicolumn{4}{|c||}{$l=2$} &\multicolumn{4}{|c|}{$l=3$}   \\
\hline
 \diagbox{$c_{2}$}{$\beta_{m}$}& 0.5 & 0.6 & 0.7 & 0.8 &  0.5 & 0.6 & 0.7 & 0.8  \\
\hline
-1/300  & $\mu\lesssim10^{-8}$&$\mu\lesssim10^{-9}$ & $\mu\lesssim10^{-10}$& &$\mu\lesssim0.08$ &$\mu\lesssim0.05$ & $\mu\lesssim10^{-2}$& $\mu\lesssim10^{-3}$\\
         &$V_{max}\sim1.9$ &$V_{max}\sim0.9$ & $V_{max}\sim0.5$& & $V_{max}\sim3.4$& $V_{max}\sim1.6$&$V_{max}\sim0.8$ &$V_{max}\sim0.4$ \\
\hline
-2/300   & $\mu\lesssim10^{-8}$&$\mu\lesssim10^{-9}$ & $\mu\lesssim10^{-10}$& &$\mu\lesssim0.08$ &$\mu\lesssim0.05$ & $\mu\lesssim10^{-2}$& $\mu\lesssim10^{-3}$\\
         &$V_{max}\sim3.9$ &$V_{max}\sim1.9$ & $V_{max}\sim0.9$& & $V_{max}\sim6.8$& $V_{max}\sim3.3$&$V_{max}\sim1.6$ &$V_{max}\sim0.8$ \\
\hline
-3/300  & $\mu\lesssim10^{-8}$&$\mu\lesssim10^{-9}$ & $\mu\lesssim10^{-10}$& &$\mu\lesssim0.08$ &$\mu\lesssim0.05$ & $\mu\lesssim10^{-2}$& $\mu\lesssim10^{-3}$\\
         &$V_{max}\sim5.9$ &$V_{max}\sim2.7$ & $V_{max}\sim1.4$& & $V_{max}\sim10.2$& $V_{max}\sim5.0$&$V_{max}\sim2.5$ &$V_{max}\sim1.2$ \\
\hline
-4/300  & & & & &$\mu\lesssim10^{-5}$ &$\mu\lesssim10^{-5}$ & $\mu\lesssim10^{-6}$& $\mu\lesssim10^{-8}$\\
         & & & & & $V_{max}\sim13.7$& $V_{max}\sim6.7$&$V_{max}\sim3.3$ &$V_{max}\sim1.5$ \\
\hline
\hline
& \multicolumn{4}{|c||}{$l=4$} &\multicolumn{4}{|c|}{$l=5$}   \\
\hline
-1/300  & $\mu\lesssim0.1$&$\mu\lesssim0.05$ & $\mu\lesssim0.01$&$\mu\lesssim10^{-3}$ &$\mu\lesssim0.4$ &$\mu\lesssim0.3$ & $\mu\lesssim0.05$& $\mu\lesssim0.004$\\
         &$V_{max}\sim5.2$ &$V_{max}\sim2.6$ & $V_{max}\sim1.3$&$V_{max}\sim0.6$ & $V_{max}\sim7.5$& $V_{max}\sim3.7$&$V_{max}\sim1.8$ &$V_{max}\sim0.9$ \\
\hline
-2/300   & $\mu\lesssim0.1$&$\mu\lesssim0.05$ & $\mu\lesssim0.01$&$\mu\lesssim10^{-3}$ &$\mu\lesssim0.2$ &$\mu\lesssim0.2$ & $\mu\lesssim0.05$& $\mu\lesssim0.003$\\
         &$V_{max}\sim10.5$ &$V_{max}\sim5.1$ & $V_{max}\sim2.5$&$V_{max}\sim1.2$ & $V_{max}\sim15.0$& $V_{max}\sim7.4$&$V_{max}\sim3.7$ &$V_{max}\sim1.7$ \\
\hline
-3/300  & $\mu\lesssim0.1$&$\mu\lesssim0.05$ & $\mu\lesssim0.01$&$\mu\lesssim10^{-3}$ &$\mu\lesssim0.2$ &$\mu\lesssim0.1$ & $\mu\lesssim0.04$& $\mu\lesssim0.003$\\
         &$V_{max}\sim15.8$ &$V_{max}\sim7.7$ & $V_{max}\sim3.8$& $V_{max}\sim1.8$ & $V_{max}\sim22.6$& $V_{max}\sim11.0$&$V_{max}\sim5.5$ &$V_{max}\sim2.6$ \\
\hline
-4/300  &$\mu\lesssim10^{-5}$ & $\mu\lesssim10^{-7}$&$\mu\lesssim10^{-7}$ & $\mu\lesssim10^{-8}$&$\mu\lesssim10^{-6}$ &$\mu\lesssim10^{-8}$ & $\mu\lesssim10^{-9}$& $\mu\lesssim10^{-10}$\\
         &$V_{max}\sim21.1$ &$V_{max}\sim10.3$ &$V_{max}\sim5.1$ &$V_{max}\sim2.4$ & $V_{max}\sim30.1$& $V_{max}\sim14.7$&$V_{max}\sim7.3$ &$V_{max}\sim3.4$ \\
\hline
\end{tabular}
\label{Tab1}
\end{table}


\begin{thebibliography}{}

\bibitem{Penrose:1964wq}
R.~Penrose,
``Gravitational collapse and space-time singularities,''
Phys. Rev. Lett. \textbf{14} (1965), 57-59
doi:10.1103/PhysRevLett.14.57

\bibitem{Akiyama:2019cqa}
K.~Akiyama \textit{et al.} [Event Horizon Telescope],
``First M87 Event Horizon Telescope Results. I. The Shadow of the Supermassive Black Hole,''
Astrophys. J. \textbf{875} (2019) no.1, L1
doi:10.3847/2041-8213/ab0ec7
[arXiv:1906.11238 [astro-ph.GA]].

\bibitem{Garofalo:2020ajg}
D.~Garofalo,
``Spin of the M87 black hole,''
Annalen Phys. \textbf{532} (2020) no.4, 1900480
doi:10.1002/andp.201900480
[arXiv:2003.02163 [astro-ph.HE]].

\bibitem{Vincent:2020dij}
F.~H.~Vincent, M.~Wielgus, M.~A.~Abramowicz, E.~Gourgoulhon, J.~P.~Lasota, T.~Paumard and G.~Perrin,
``Geometric modeling of M87* as a Kerr black hole or a non-Kerr compact object,''
[arXiv:2002.09226 [gr-qc]].

\bibitem{Dokuchaev:2020rye}
V.~I.~Dokuchaev and N.~O.~Nazarova,
``Modeling the motion of a bright spot in jets from black holes M87* and SgrA*,''
[arXiv:2010.01885 [astro-ph.HE]].

\bibitem{Stepanian:2021vvk}
A.~Stepanian, S.~Khlghatyan and V.~G.~Gurzadyan,
``Black hole shadow to probe modified gravity,''
Eur. Phys. J. Plus \textbf{136} (2021) no.1, 127
doi:10.1140/epjp/s13360-021-01119-2
[arXiv:2101.08261 [gr-qc]].

\bibitem{TheLIGOScientific:2016src}
B.~P.~Abbott \textit{et al.} [LIGO Scientific and Virgo],
``Tests of general relativity with GW150914,''
Phys. Rev. Lett. \textbf{116} (2016) no.22, 221101
[erratum: Phys. Rev. Lett. \textbf{121} (2018) no.12, 129902]
doi:10.1103/PhysRevLett.116.221101
[arXiv:1602.03841 [gr-qc]].

\bibitem{Cornish:2017jml}
N.~Cornish, D.~Blas and G.~Nardini,
``Bounding the speed of gravity with gravitational wave observations,''
Phys. Rev. Lett. \textbf{119} (2017) no.16, 161102
doi:10.1103/PhysRevLett.119.161102
[arXiv:1707.06101 [gr-qc]].

\bibitem{Hawking:1974sw}
S.~W.~Hawking,
``Particle Creation by Black Holes,''
Commun. Math. Phys. \textbf{43} (1975), 199-220
[erratum: Commun. Math. Phys. \textbf{46} (1976), 206]
doi:10.1007/BF02345020

\bibitem{Hawking:1976de}
S.~W.~Hawking,
``Black Holes and Thermodynamics,''
Phys. Rev. D \textbf{13} (1976), 191-197
doi:10.1103/PhysRevD.13.191


\bibitem{Fernando:2004ay}
S.~Fernando,
``Greybody factors of charged dilaton black holes in 2 + 1 dimensions,''
Gen. Rel. Grav. \textbf{37} (2005), 461-481
doi:10.1007/s10714-005-0035-x
[arXiv:hep-th/0407163 [hep-th]].

\bibitem{Ahmed:2016lou}
J.~Ahmed and K.~Saifullah,
``Greybody factor of a scalar field from Reissner\textendash{}Nordstr\"om\textendash{}de Sitter black hole,''
Eur. Phys. J. C \textbf{78} (2018) no.4, 316
doi:10.1140/epjc/s10052-018-5800-6
[arXiv:1610.06104 [gr-qc]].

\bibitem{Sharif:2019yiy}
M.~Sharif and Q.~Ama-Tul-Mughani,
``Greybody Factor for a Rotating Bardeen Black Hole,''
Eur. Phys. J. Plus \textbf{134} (2019) no.12, 616
doi:10.1140/epjp/i2019-12979-0
[arXiv:1909.02862 [gr-qc]].

\bibitem{Sharif:2020hyz}
M.~Sharif and Q.~Ama-Tul-Mughani,
``Greybody factor for quintessential Kerr\textendash{}Newman black hole,''
Phys. Dark Univ. \textbf{27} (2020), 100436
doi:10.1016/j.dark.2019.100436
[arXiv:2001.10798 [gr-qc]].


\bibitem{Parikh:1999mf}
M.~K.~Parikh and F.~Wilczek,
``Hawking radiation as tunneling,''
Phys. Rev. Lett. \textbf{85} (2000), 5042-5045
doi:10.1103/PhysRevLett.85.5042
[arXiv:hep-th/9907001 [hep-th]].

\bibitem{Cho:2004wj}
H.~T.~Cho and Y.~C.~Lin,
``WKB analysis of the scattering of massive Dirac fields in Schwarzschild black hole spacetimes,''
Class. Quant. Grav. \textbf{22} (2005), 775-790
doi:10.1088/0264-9381/22/5/001
[arXiv:gr-qc/0411090 [gr-qc]].

\bibitem{Konoplya:2010kv}
R.~A.~Konoplya and A.~Zhidenko,
``Passage of radiation through wormholes of arbitrary shape,''
Phys. Rev. D \textbf{81} (2010), 124036
doi:10.1103/PhysRevD.81.124036
[arXiv:1004.1284 [hep-th]].

\bibitem{Dey:2018cws}
S.~Dey and S.~Chakrabarti,
``A note on electromagnetic and gravitational perturbations of the Bardeen de Sitter black hole: quasinormal modes and greybody factors,''
Eur. Phys. J. C \textbf{79} (2019) no.6, 504
doi:10.1140/epjc/s10052-019-7004-0
[arXiv:1807.09065 [gr-qc]].

\bibitem{Konoplya:2019ppy}
R.~A.~Konoplya and A.~F.~Zinhailo,
``Hawking radiation of non-Schwarzschild black holes in higher derivative gravity: a crucial role of grey-body factors,''
Phys. Rev. D \textbf{99} (2019) no.10, 104060
doi:10.1103/PhysRevD.99.104060
[arXiv:1904.05341 [gr-qc]].

\bibitem{Konoplya:2019hlu}
R.~A.~Konoplya, A.~Zhidenko and A.~F.~Zinhailo,
``Higher order WKB formula for quasinormal modes and grey-body factors: recipes for quick and accurate calculations,''
Class. Quant. Grav. \textbf{36} (2019), 155002
doi:10.1088/1361-6382/ab2e25
[arXiv:1904.10333 [gr-qc]].

\bibitem{Devi:2020uac}
S.~Devi, R.~Roy and S.~Chakrabarti,
``Quasinormal modes and greybody factors of the novel four dimensional Gauss\textendash{}Bonnet black holes in asymptotically de Sitter space time: scalar, electromagnetic and Dirac perturbations,''
Eur. Phys. J. C \textbf{80} (2020) no.8, 760
doi:10.1140/epjc/s10052-020-8311-1
[arXiv:2004.14935 [gr-qc]].


\bibitem{Visser:1998ke}
M.~Visser,
``Some general bounds for 1-D scattering,''
Phys. Rev. A \textbf{59} (1999), 427-438
doi:10.1103/PhysRevA.59.427
[arXiv:quant-ph/9901030 [quant-ph]].

\bibitem{Boonserm:2008qf}
P.~Boonserm and M.~Visser,
``Bounding the Bogoliubov coefficients,''
Annals Phys. \textbf{323} (2008), 2779-2798
doi:10.1016/j.aop.2008.02.002
[arXiv:0801.0610 [quant-ph]].

\bibitem{Boonserm:2009zba}
P.~Boonserm,
``Rigorous bounds on Transmission, Reflection, and Bogoliubov coefficients,''
[arXiv:0907.0045 [math-ph]].

\bibitem{Boonserm:2017qcq}
P.~Boonserm, T.~Ngampitipan and P.~Wongjun,
``Greybody factor for black holes in dRGT massive gravity,''
Eur. Phys. J. C \textbf{78} (2018) no.6, 492
doi:10.1140/epjc/s10052-018-5975-x
[arXiv:1705.03278 [gr-qc]].

\bibitem{Boonserm:2019mon}
P.~Boonserm, T.~Ngampitipan and P.~Wongjun,
``Greybody factor for black string in dRGT massive gravity,''
Eur. Phys. J. C \textbf{79} (2019) no.4, 330
doi:10.1140/epjc/s10052-019-6827-z
[arXiv:1902.05215 [gr-qc]].

\bibitem{Barman:2019vst}
S.~Barman,
``The Hawking effect and the bounds on greybody factor for higher dimensional Schwarzschild black holes,''
Eur. Phys. J. C \textbf{80} (2020) no.1, 50
doi:10.1140/epjc/s10052-020-7613-7
[arXiv:1907.09228 [gr-qc]].

\bibitem{Chowdhury:2020bdi}
A.~Chowdhury and N.~Banerjee,
``Greybody factor and sparsity of Hawking radiation from a charged spherical black hole with scalar hair,''
Phys. Lett. B \textbf{805} (2020), 135417
doi:10.1016/j.physletb.2020.135417
[arXiv:2002.03630 [gr-qc]].



\bibitem{Kanzi:2020cyv}
S.~Kanzi, S.~H.~Mazharimousavi and \.I.~Sakall\i{},
``Greybody factors of black holes in dRGT massive gravity coupled with nonlinear electrodynamics,''
Annals Phys. \textbf{422} (2020), 168301
doi:10.1016/j.aop.2020.168301
[arXiv:2007.05814 [hep-th]].

\bibitem{Supernova}
Supernova Search Team Collaboration, A. G. Riess et al., Astron. J. \textbf{116}, 1009-1038 (1998), [arXiv:astro-ph/9805201].

\bibitem{Supernova2}
Supernova Cosmology Project Collaboration, S. Perlmutter et al., Astrophys. J. \textbf{517}, 565-586 (1999), [arXiv:astro-ph/9812133].

\bibitem{deRham:2010ik}
C.~de Rham and G.~Gabadadze,
Generalization of the Fierz-Pauli Action,
Phys.\ Rev.\  {\bf D82}, 044020 (2010)
  [arXiv:1007.0443 [hep-th]].

\bibitem{deRham:2010kj}
C.~de Rham, G.~Gabadadze and A.~J.~Tolley,
Resummation of Massive Gravity,
Phys.\ Rev.\ Lett.\  {\bf 106}, 231101 (2011)
  [arXiv:1011.1232].

 \bibitem{Berezhiani:2011mt}
L.~Berezhiani, G.~Chkareuli, C.~de Rham, G.~Gabadadze and A.~J.~Tolley,
On Black Holes in Massive Gravity,
Phys.\ Rev.\ D {\bf 85}, 044024 (2012)
  doi:10.1103/PhysRevD.85.044024
  [arXiv:1111.3613 [hep-th]].

\bibitem{Brito:2013xaa}
R.~Brito, V.~Cardoso and P.~Pani,
Black holes with massive graviton hair,
Phys.\ Rev.\ D {\bf 88}, 064006 (2013)
  doi:10.1103/PhysRevD.88.064006
  [arXiv:1309.0818 [gr-qc]].

\bibitem{Volkov:2013roa}
M.~S.~Volkov,
Self-accelerating cosmologies and hairy black holes in ghost-free bigravity and massive gravity,
Class.\ Quant.\ Grav.\  {\bf 30}, 184009 (2013)

\bibitem{Cai:2012db}
Y.~F.~Cai, D.~A.~Easson, C.~Gao and E.~N.~Saridakis,
Charged black holes in nonlinear massive gravity,
Phys.\ Rev.\ D {\bf 87}, 064001 (2013)
  doi:10.1103/PhysRevD.87.064001
  [arXiv:1211.0563 [hep-th]].

\bibitem{Babichev:2014fka}
E.~Babichev and A.~Fabbri,
A class of charged black hole solutions in massive (bi)gravity,
JHEP {\bf 1407}, 016 (2014)
  doi:10.1007/JHEP07(2014)016
  [arXiv:1405.0581 [gr-qc]].

\bibitem{Babichev:2015xha}
E.~Babichev and R.~Brito,
Black holes in massive gravity,
Class.\ Quant.\ Grav.\  {\bf 32}, 154001 (2015)
  doi:10.1088/0264-9381/32/15/154001
  [arXiv:1503.07529 [gr-qc]].

\bibitem{Hu:2016hpm}
Y.~P.~Hu, X.~M.~Wu and H.~Zhang,
Generalized Vaidya Solutions and Misner-Sharp mass for $n$-dimensional massive gravity,
Phys.\ Rev.\ D {\bf 95}, no. 8, 084002 (2017)
  doi:10.1103/PhysRevD.95.084002
  [arXiv:1611.09042 [gr-qc]].


\bibitem{Cai:2014znn}
R.~G.~Cai, Y.~P.~Hu, Q.~Y.~Pan and Y.~L.~Zhang,
Thermodynamics of Black Holes in Massive Gravity,
Phys.\ Rev.\ D {\bf 91}, no. 2, 024032 (2015)
  doi:10.1103/PhysRevD.91.024032
  [arXiv:1409.2369 [hep-th]].

\bibitem{Ghosh:2015cva}
S.~G.~Ghosh, L.~Tannukij and P.~Wongjun,
A class of black holes in dRGT massive gravity and their thermodynamical properties,
Eur.\ Phys.\ J.\ C {\bf 76}, no. 3, 119 (2016)
  doi:10.1140/epjc/s10052-016-3943-x
  [arXiv:1506.07119 [gr-qc]].

\bibitem{Hinterbichler}
K. Hinterbichler, ``Theoretical aspects of massive gravity", {\it Reviews of Modern Physics} \textbf{84}, 671-710, 2012, [arXiv: 1105.3735 [hep-th]].

\bibitem{deRham:2014zqa}
  C.~de Rham,
  ``Massive Gravity,''
  Living Rev.\ Rel.\  {\bf 17}, 7 (2014)
  doi:10.12942/lrr-2014-7
  [arXiv:1401.4173 [hep-th]].
  
\bibitem{Hou:2020yni}
M.~S.~Hou, H.~Xu and Y.~C.~Ong,
``Hawking Evaporation of Black Holes in Massive Gravity,''
Eur. Phys. J. C \textbf{80} (2020) no.11, 1090
doi:10.1140/epjc/s10052-020-08678-1
[arXiv:2008.10049 [hep-th]].


\bibitem{Iyer:1986np}
S.~Iyer and C.~M.~Will,
``Black Hole Normal Modes: A \{WKB\} Approach. 1. Foundations and Application of a Higher Order \{WKB\} Analysis of Potential Barrier Scattering,''
Phys. Rev. D \textbf{35} (1987), 3621
doi:10.1103/PhysRevD.35.3621

\bibitem{Simone:1991wn}
L.~E.~Simone and C.~M.~Will,
``Massive scalar quasinormal modes of Schwarzschild and Kerr black holes,''
Class. Quant. Grav. \textbf{9} (1992), 963-978
doi:10.1088/0264-9381/9/4/012

\bibitem{Cho:2003qe}
H.~T.~Cho,
``Dirac quasinormal modes in Schwarzschild black hole space-times,''
Phys. Rev. D \textbf{68} (2003), 024003
doi:10.1103/PhysRevD.68.024003
[arXiv:gr-qc/0303078 [gr-qc]].

\bibitem{Konoplya:2003ii}
R.~A.~Konoplya,
``Quasinormal behavior of the d-dimensional Schwarzschild black hole and higher order WKB approach,''
Phys. Rev. D \textbf{68} (2003), 024018
doi:10.1103/PhysRevD.68.024018
[arXiv:gr-qc/0303052 [gr-qc]].

\bibitem{Chen:2019kaq}
P.~Wongjun, C.~H.~Chen and R.~Nakarachinda,
``Quasinormal modes of a massless Dirac field in de Rham-Gabadadze-Tolley massive gravity,''
Phys. Rev. D \textbf{101} (2020) no.12, 124033
doi:10.1103/PhysRevD.101.124033
[arXiv:1910.05908 [gr-qc]].

\end{thebibliography}
\end{document}